\documentclass[fleqn,usenatbib]{mnras}

\usepackage{newtxtext,newtxmath}

\usepackage[T1]{fontenc}

\DeclareRobustCommand{\VAN}[3]{#2}
\let\VANthebibliography\thebibliography
\def\thebibliography{\DeclareRobustCommand{\VAN}[3]{##3}\VANthebibliography}

\usepackage{graphicx}	
\usepackage{amsmath}	
\usepackage{xspace}  
\usepackage{amsmath}
\usepackage{CJKutf8}
\usepackage{multirow}  
\usepackage{tabularx}
\usepackage{makecell}
\usepackage{supertabular,booktabs}
\renewcommand{\toprule}{\hline\hline}
\renewcommand{\midrule}{\hline}
\renewcommand{\bottomrule}{\hline}

\newcommand{\mASIAA}{Institute of Astronomy and Astrophysics, Academia Sinica, No. 1, Sec. 4, Roosevelt Road, Taipei 10617, Taiwan}

\newcommand{\mCASS}{Center for Astrophysics and Space Sciences, Department of Physics, University of California, San Diego, 9500 Gilman Drive, La Jolla, CA 92093, USA}

\newcommand{\mMPIA}{Max-Planck-Institut f\"ur Astronomie, K\"onigstuhl 17, D-69117 Heidelberg, Germany}

\newcommand{\mOSUa}{Department of Astronomy, The Ohio State University, 140 West 18th Avenue, Columbus, OH 43210, USA}

\newcommand{\mCCAPP}{Center for Cosmology and Astroparticle Physics, 191 West Woodruff Avenue, Columbus, OH 43210, USA}

\newcommand{\mUOA}{Department of Physics, University of Alberta, 4-183 CCIS, Edmonton, AB T6G 2E1, Canada}

\newcommand{\mUGENT}{Sterrenkundig Observatorium, Universiteit Gent, Krijgslaan 281 S9, B-9000 Gent, Belgium}

\newcommand{\mMcMaster}{Department of Physics and Astronomy, McMaster University, 1280 Main Street West, Hamilton, ON L8S 4M1, Canada}

\newcommand{\mCITA}{Canadian Institute for Theoretical Astrophysics (CITA), University of Toronto, 60 St George Street, Toronto, ON M5S 3H8, Canada}

\newcommand{\GHz}{{\rm GHz}\xspace}
\newcommand{\SigmaMassUnit}{\ensuremath{\rm M_\odot~pc^{-2}}\xspace}
\newcommand{\SigmaMassUnitKpc}{\ensuremath{\rm M_\odot\,kpc\textsuperscript{-2}}\xspace}
\newcommand{\SigmasfrUnit}{\ensuremath{{\rm M_\odot\,yr}^{-1}\,{\rm kpc}^{-2}}\xspace}

\newcommand{\ICOUnit}{{\rm K\,km\,s\textsuperscript{-1}}\xspace}
\newcommand{\SEDUnit}{{\rm MJy\,sr\textsuperscript{-1}}\xspace}
\newcommand{\acoUnit}{\ensuremath{{\rm M_\odot\,pc^{-2}\,(K\,km\,s^{-1})^{-1}}}\xspace}
\newcommand{\XcoUnit}{\ensuremath{\rm cm^{-2}\,(K\,km\,s^{-1})^{-1}}\xspace}

\newcommand{\metal}{12+\logt({\rm O/H})\xspace}
\newcommand{\CO}{{\rm CO}\xspace}

\newcommand{\HI}{\textsc{H\,i}\xspace}
\newcommand{\HII}{\textsc{H\,ii}\xspace}
\newcommand{\HTWO}{{\rm H}\textsubscript{2}\xspace}

\newcommand{\Tclr}{\ensuremath{{T_\mathrm{clr}}}\xspace}
\newcommand{\Td}{\ensuremath{{T_\mathrm{dust}}}\xspace}
\newcommand{\Tumin}{\ensuremath{{T_\mathrm{dust}(U_\mathrm{min})}}\xspace}
\newcommand{\Tubar}{\ensuremath{{T_\mathrm{dust}(\overline{U})}}\xspace}

\newcommand{\Sigmad}{\ensuremath{\Sigma_\mathrm{dust}}\xspace}
\newcommand{\Sigmastar}{\ensuremath{\Sigma_\star}\xspace}
\newcommand{\Sigmasfr}{\ensuremath{\Sigma_\mathrm{SFR}}\xspace}
\newcommand{\Sigmagas}{\ensuremath{\Sigma_\mathrm{gas}}\xspace}

\newcommand{\Sigmaatom}{\ensuremath{\Sigma_\mathrm{atom}}\xspace}
\newcommand{\Sigmamol}{\ensuremath{\Sigma_\mathrm{mol}}\xspace}
\newcommand{\Sigmatot}{\ensuremath{\Sigma_\mathrm{Total}}\xspace}

\newcommand{\aco}{\ensuremath{\alpha_\CO}\xspace}
\newcommand{\acoMW}{\ensuremath{\alpha_\CO^{\rm MW}}\xspace}
\newcommand{\acoAccurso}{\ensuremath{\alpha_\CO^{\rm A17}}\xspace}
\newcommand{\acoBolatto}{\ensuremath{\alpha_\CO^{\rm B13}}\xspace}

\newcommand{\Ubar}{\ensuremath{\overline{U}}\xspace}
\newcommand{\Umin}{\ensuremath{U_\mathrm{min}}\xspace}

\usepackage{threeparttable}
\usepackage[pagewise]{lineno}
\newcommand\arcdeg{\mbox{$^\circ$}}%
\newcommand\nodata{\mbox{-{}-}}%
\renewcommand{\metal}{12+\ensuremath{\log(\mathrm{ O/H})}\xspace}

\defcitealias{Hirashita_Chiang22}{Paper~I}

\title[Kpc-scale properties of dust temperature]{
Kpc-scale properties of dust temperature in terms of dust mass and star formation activity
}

\author[I-D. Chiang et al.]{
\parbox[t]{\textwidth}{
I-Da Chiang (\begin{CJK*}{UTF8}{bkai}江宜達\end{CJK*})$^{1}$\thanks{E-mail: idchiang@asiaa.sinica.edu.tw}, 
Hiroyuki Hirashita$^{1,2}$,  
J\'er\'emy Chastenet$^{3}$,  
Eric~W.~Koch$^{4}$, 
Adam K. Leroy$^{5,6}$, 
Erik Rosolowsky$^{7}$, 
Karin M. Sandstrom$^{8}$, 
Amy Sardone$^{5,6}$,  
Jiayi Sun \begin{CJK*}{UTF8}{gbsn}(孙嘉懿)\end{CJK*}$^{9,10}$, 
Thomas G. Williams$^{11, 12}$  
}\\
$^{1}$\mASIAA\\
$^{2}$Theoretical Astrophysics, Department of Earth and Space Science, Osaka University,
1-1 Machikaneyama, Toyonaka, Osaka 560-0043, Japan\\
$^{3}$\mUGENT\\
$^{4}$Center for Astrophysics $|$ Harvard \& Smithsonian, 60 Garden Street, Cambridge, MA 02138, USA\\
$^{5}$\mOSUa\\
$^{6}$\mCCAPP\\
$^{7}$\mUOA\\
$^{8}$\mCASS\\
$^{9}$\mMcMaster\\
$^{10}$\mCITA\\
$^{11}$\mMPIA\\
$^{12}$Sub-department of Astrophysics, Department of Physics, University of Oxford, Keble Road, Oxford OX1 3RH, UK
}

\date{Accepted XXX. Received YYY; in original form ZZZ}

\pubyear{2022}

\begin{document}
\label{firstpage}
\pagerange{\pageref{firstpage}--\pageref{lastpage}}
\maketitle

\begin{abstract}

We investigate how the dust temperature is affected by local environmental quantities, especially dust surface density ($\Sigma_\mathrm{dust}$), dust-to-gas ratio (D/G) and interstellar radiation field.
We compile multi-wavelength observations in 46 nearby galaxies, uniformly processed with a common physical resolution of 2~kpc. A physical dust model is used to fit the infrared dust emission spectral energy distribution (SED) observed with \textit{WISE} and \textit{Herschel}. The star formation rate (SFR) is traced with \textit{GALEX} ultraviolet data corrected by \textit{WISE} infrared.
We find that the dust temperature correlates well with the SFR surface density ($\Sigma_{\rm SFR}$), which traces the radiation from young stars. The dust temperature decreases with increasing D/G at fixed $\Sigma_{\rm SFR}$ as expected from stronger dust shielding at high D/G, when $\Sigma_\mathrm{SFR}$ is higher than $\sim 2\times 10^{-3}~\rm M_\odot~yr^{-1}~kpc^{-2}$. These measurements are in good agreement with the dust temperature predicted by our proposed analytical model.
Below this range of $\Sigma_\mathrm{SFR}$, the observed dust temperature is higher than the model prediction and is only weakly dependent on D/G, which is possibly due to the dust heating from old stellar population or the variation of SFR within the past $10^{10}$~yr.
Overall, the dust temperature as a function of $\Sigma_\mathrm{SFR}$ and $\Sigma_\mathrm{dust}$ predicted by our analytical model is consistent with observations.
We also notice that at fixed gas surface density, $\Sigma_{\rm SFR}$ tends to increase with D/G, i.e. we can empirically modify the Kennicutt-Schmidt law with a dependence on D/G to better match observations.
\end{abstract}




\section{Introduction}\label{sec:intro}
Dust is a key component in the interstellar medium (ISM). In the diffuse ISM of the Milky Way (MW), roughly 20--50 per cent of metals reside in solid dust grains according to elemental depletions \citep[$F_*=0$ to 1 in][]{JENKINS09}. Dust grains absorb and scatter a significant fraction of starlight in galaxies \citep[e.g.\ 30 per cent suggested in][]{BERNSTEIN02}, and re-radiates the absorbed energy in the infrared \citep[IR,][]{CALZETTI01, BUAT12}.
These processes regulate the spectral energy distribution (SED) of the interstellar radiation field (ISRF). Dust catalyzes the formation of \HTWO \citep{GOULD63, DRAINE03, CAZAUX04, YAMASAWA11, GALLIANO18}, which helps the formation of the molecular phase ISM.
Because of the major roles of dust in galaxy evolution and chemistry, it is important to track the interaction between dust and the ISRF in various environments in the ISM. In particular, the dust surface density is of critical importance because it is not only an indicator of the dust mass but also is proportional to the dust optical depth.

The dust surface density (\Sigmad) in the ISM can be obtained by analyzing the dust emission SED in the IR \citep{DESERT90,DRAINE07,COMPIEGNE11,JONES17_THEMIS,Relano20,HensleyDraine22}. The intensity of dust emission is proportional to the dust mass at fixed wavelength and is a monotonically increasing function of the dust temperature. 
The dust temperature (\Td), or the radiation field heating up dust in some models, can be constrained with the ratio between dust emission at different photometric bands in the measured SED. With the obtained value for \Td, we are able to derive \Sigmad from the intensity of dust emission with an assumption of grain emissivity.

With multi-band far-IR (FIR) data, especially the 70--500~\micron\ photometry obtained with the \textit{Herschel Space Observatory} \citep{PILBRATT10}, there have been many efforts to measure spatially resolved, emission-based \Sigmad and \Td in nearby galaxies \citep[e.g.][]{ANIANO12,Bendo12a,SMITH12,DRAINE14,GORDON14,Tabatabaei14,Bendo15,Davies17_DustPedia,Bianchi18,UTOMO19,Vilchez19,ANIANO20,Chiang21,Nersesian21}. Meanwhile, observations of distant (high-redshift) galaxies are often performed with the Atacama Large Millimeter/submillimeter Array (ALMA) because of increased requirements for spatial resolution and sensitivity \citep[e.g.][]{Capak15,Watson15,Bouwens16,Liu19_A3COSMOS1}. 
Due to the narrow instantaneous frequency coverage of ALMA, each galaxy is observed in a limited number of wavelength bands, usually only 1--2 bands, and as a result the dust temperatures in these systems remain comparatively uncertain.
In spite of these limitations, the observed dust temperatures by ALMA have provided meaningful insight into the physical conditions in the ISM at high redshift. Studies that succeeded in obtaining multi-band ALMA data suggested that high-redshift \Td spans $\sim$30--70~K \citep[see the $z$$\sim$5--8.3 data compiled in][]{Burgarella20,Bakx21}, which is not only warmer but also showing a larger variety than the dust temperatures observed in nearby disc galaxies, e.g.\ $\sim$17--40~K in the KINGFISH samples \citep{Dale12} and $\sim$15--25~K in the Local Group  \citep{UTOMO19}. High dust temperatures in distant galaxies are also indicated by some indirect estimates using the ultraviolet (UV) optical depth or the [\textsc{C~ii}] 158~\micron\ line \citep{Sommovigo22,Ferrara22}.
These results indicate that the ISRFs are high probably because of intense star formation activity in high-redshift galaxies.

In \citet[][hereafter \citetalias{Hirashita_Chiang22}]{Hirashita_Chiang22}, we built analytical models that evaluate how dust temperature varies with relevant physical quantities, i.e.\ dust surface density and ISRF, to clarify the physical conditions that possibly cause the observed high and scattered \Td in high-redshift galaxies. With these models, we predicted that to the first order, \Td increases with UV radiation from stars (traced by the star formation rate surface density, \Sigmasfr). This supports the above-mentioned view of intense star formation activity in high-redshift galaxies. We also found that, at fixed \Sigmasfr, dust temperature increases towards lower dust-to-gas ratio (D/G).
This is because of less shielding of ISRF by dust \citep[see also][]{Sommovigo22}. Moreover, with \Sigmasfr and \Sigmad measured, we can roughly predict the dust temperature, which offers a way to improve the dust mass and temperature estimation iteratively. We tested these predictions against several high-redshift observations.

Although we obtained useful insights using our model, the large uncertainties and low spatial resolutions of high-redshift observations hampered rigid conclusions. Observations with greater precision and larger sample size are useful to validate our dust temperature model in \citetalias{Hirashita_Chiang22}.
For this purpose, observations in the nearby galaxies are suitable in terms of both precision and sample size, as demonstrated by the multi-wavelength datasets in the literature \citep[e.g.][]{Barrera-Ballesteros16,Barrera-Ballesteros21_EDGE_CALIFA,Casasola17,Casasola22,LEROY19,Enia20,Morselli20,Ellison21_ALMaQUEST_V,Sanchez21_EDGE_CALIFA,Abdurro'uf22_SDSS,Sun22_MegaTable}.
These observations also provide measurements at the low \Sigmasfr and \Sigmad end, which was not tested in \citetalias{Hirashita_Chiang22}. With the availability of multi-band FIR photometric data, e.g.\ observations made with \textit{Herschel}, and gas surface densities from high sensitivity emission line observations, e.g.\ The \HI Nearby Galaxy Survey (THINGS) \citep{WALTER08}, The HERA CO-Line Extragalactic Survey (HERACLES) \citep{LEROY09} and the ALMA, VLA and MeerKAT surveys conducted in the Physics at High Angular resolution in Nearby Galaxies (PHANGS) project\footnote{\url{http://phangs.org/}} \citep{Leroy21_PHANGS-ALMA_CO,Sun22_MegaTable}, we are able to derive dust temperature and D/G more robustly for nearby galaxies.

In this work, we compare our model to a sample of nearby galaxies to investigate what regulates the dust temperature. We compile spatially resolved, multi-wavelength data for 46 nearby galaxies. We uniformly process them with a 2~kpc resolution. We derive key quantities that affect the dust temperature with the following methods: \Sigmad is derived from the dust emission SED observed with the \textit{Wide-field Infrared Survey Explorer} \citep[\textit{WISE},][]{WRIGHT10} and \textit{Herschel} using the \citet{DRAINE07} physical dust model including the correction factor introduced by \citet{Chastenet21_M101}.
We utilize the \textit{Galaxy Evolution Explorer} \citep[\textit{GALEX},][]{MARTIN05} UV data, supplemented by \textit{WISE} IR, to trace star formation rate (SFR) following the \citet{LEROY19} prescription.
Besides dust temperature, we also examine the scaling relations among the surface densities of dust, SFR and gas, which are necessary to setup the \citetalias{Hirashita_Chiang22} model.
These measurements and comparisons allow us to test the model in \citetalias{Hirashita_Chiang22} and to clarify how key physical properties, especially star formation activity and dust surface density, regulate the dust temperature.

This paper is organized as follows.
In Section~\ref{sec:data}, we introduce the sample galaxies and data sets used in this work.
In Section~\ref{sec:method}, we describe how we uniformly process the multi-wavelength observations and convert them to desired physical quantities. We also outline how we adapt the \citetalias{Hirashita_Chiang22} model in this work.
We present our measured dust temperature and relevant scaling relations in Section~\ref{sec:measurements}.
We compare our measurements to the \citetalias{Hirashita_Chiang22} model and provide some extended discussions in Section~\ref{sec:comparison}.
Finally, we summarize our key findings in Section~\ref{sec:summary}.

\section{Sample and Data}\label{sec:data}
The data necessary for this study are dust surface densities and temperatures from IR photometry, gas surface densities from CO and \HI emission lines and SFRs from UV and IR photometry. 
We select our sample galaxies from the $z=0$ Multiwavelength Galaxy Synthesis ($z$0MGS) catalog \citep[][J. Chastenet et al. in preparation]{LEROY19}. We pick the galaxies with \textit{Herschel} IR, \textit{WISE} IR and \textit{GALEX} UV data available as the master sample.
From this large sample, we pick 49 galaxies with both low-$J$ CO rotational lines and \HI data from archival or our new data.
We set the desired physical resolution at 2~kpc, which puts a limitation on distance at $ D \lesssim 20~\mathrm{Mpc}$, corresponding to $\sim$2~kpc resolution of the coarsest resolution data obtained by the \textit{Herschel} SPIRE 250 $\micron$ band.
Finally, we exclude high-inclination ($>80^\circ$) galaxies, which yields 46 galaxies in the end.\footnote{An early version of this data set was first compiled in \citet{Chiang21PhD}.} We list the properties and data sources of these galaxies in Table~\ref{tab:samples}.

\begin{table*}
\centering
\caption{Sample galaxies.}
\label{tab:samples}
\begin{tabular}{ccccccccccc}
\toprule
  Galaxy & Dist. &       $i$ &      P.A. & $R_{25}$\hspace{0em} & $R_\mathrm{eff}$\hspace{0em} & $\log(M_\star)$ & Type &                         CO Ref &         \HI Ref & 12+log(O/H) Ref \\
         & [Mpc] & [\arcdeg] & [\arcdeg] &                [kpc] &             [kpc] &      [M$_\odot$] &      &                                &                         &                 \\
     (1) &   (2) &       (3) &       (4) &                  (5) &               (6) &             (7) &  (8) &                            (9) &                    (10) &            (11) \\
\midrule
  IC0342 &   3.4 &      31.0 &      42.0 &                  9.9 &               4.3 &            10.2 &    5 &   $a.$ &            $h.$ &            $p.$ \\
 NGC0224 &   0.8 &      77.7 &      38.0 &                 21.2 &           \nodata &         \nodata &    3 &   $b.$ &            $i.$ &            $r.$ \\
 NGC0253 &   3.7 &      75.0 &      52.5 &                 14.4 &               4.7 &            10.5 &    5 &   $c.$ &            $j.$ &         \nodata \\
 NGC0300 &   2.1 &      39.8 &     114.3 &                  5.9 &               2.0 &             9.3 &    6 &   $c.$ &            $j.$ &            $r.$ \\
 NGC0337 &  19.5 &      51.0 &      90.0 &                  8.3 &               2.4 &             9.7 &    6 &   $d.$ &            $k.$ &         \nodata \\
 NGC0598 &   0.9 &      55.0 &     201.0 &                  8.5 &               2.5 &             9.4 &    5 &   $e.$ &            $l.$ &            $r.$ \\
 NGC0628 &   9.8 &       8.9 &      20.7 &                 14.1 &               3.9 &            10.2 &    5 &   $c.$ &            $m.$ &            $q.$ \\
 NGC0925 &   9.2 &      66.0 &     287.0 &                 14.3 &               4.5 &             9.8 &    6 &   $d.$ &            $m.$ &            $r.$ \\
 NGC2403 &   3.2 &      63.0 &     124.0 &                  9.3 &               2.4 &             9.6 &    5 &   $d.$ &            $m.$ &            $r.$ \\
 NGC2841 &  14.1 &      74.0 &     153.0 &                 14.2 &               5.4 &            10.9 &    3 &   $d.$ &            $m.$ &         \nodata \\
 NGC2976 &   3.6 &      65.0 &     335.0 &                  3.0 &               1.3 &             9.1 &    5 &   $d.$ &            $m.$ &         \nodata \\
 NGC3184 &  12.6 &      16.0 &     179.0 &                 13.5 &               5.3 &            10.3 &    5 &   $d.$ &            $m.$ &            $r.$ \\
 NGC3198 &  13.8 &      72.0 &     215.0 &                 13.0 &               5.0 &            10.0 &    5 &   $d.$ &            $m.$ &         \nodata \\
 NGC3351 &  10.0 &      45.1 &     193.2 &                 10.5 &               3.0 &            10.3 &    3 &   $c.$ &            $m.$ &            $q.$ \\
 NGC3521 &  13.2 &      68.8 &     343.0 &                 16.0 &               3.9 &            11.0 &    3 &   $c.$ &            $m.$ &         \nodata \\
 NGC3596 &  11.3 &      25.1 &      78.4 &                  6.0 &               1.6 &             9.5 &    5 &   $c.$ &            $n.$ &         \nodata \\
 NGC3621 &   7.1 &      65.8 &     343.8 &                  9.8 &               2.7 &            10.0 &    6 &   $c.$ &            $m.$ &            $r.$ \\
 NGC3627 &  11.3 &      57.3 &     173.1 &                 16.9 &               3.6 &            10.7 &    3 &   $c.$ &            $m.$ &            $q.$ \\
 NGC3631 &  18.0 &      32.4 &     -65.6 &                  9.7 &               2.9 &            10.2 &    5 &   $f.$ &            $h.$ &         \nodata \\
 NGC3938 &  17.1 &      14.0 &     195.0 &                 13.4 &               3.7 &            10.3 &    5 &   $d.$ &            $k.$ &         \nodata \\
 NGC3953 &  17.1 &      61.5 &      12.5 &                 15.2 &               5.3 &            10.6 &    4 &   $f.$ &            $h.$ &         \nodata \\
 NGC4030 &  19.0 &      27.4 &      28.7 &                 10.5 &               2.1 &            10.6 &    4 &   $f.$ &            $h.$ &         \nodata \\
 NGC4051 &  17.1 &      43.4 &     -54.8 &                 14.7 &               3.7 &            10.3 &    3 &   $f.$ &            $h.$ &         \nodata \\
 NGC4207 &  15.8 &      64.5 &     121.9 &                  3.4 &               1.4 &             9.6 &    7 &   $c.$ &            $n.$ &         \nodata \\
 NGC4254 &  13.1 &      34.4 &      68.1 &                  9.6 &               2.4 &            10.3 &    5 &   $c.$ &            $k.$ &            $q.$ \\
 NGC4258 &   7.6 &      68.3 &     150.0 &                 18.7 &               5.8 &            10.7 &    4 &   $g.$ &            $o.$ &            $r.$ \\
 NGC4321 &  15.2 &      38.5 &     156.2 &                 13.5 &               5.5 &            10.7 &    3 &   $c.$ &            $k.$ &            $q.$ \\
 NGC4450 &  16.8 &      48.5 &      -6.3 &                 13.3 &               4.3 &            10.7 &    2 &   $f.$ &            $h.$ &         \nodata \\
NGC4496A &  14.9 &      53.8 &      51.1 &                  7.3 &               3.0 &             9.6 &    6 &   $c.$ &            $h.$ &         \nodata \\
 NGC4501 &  16.8 &      60.1 &     -37.8 &                 21.1 &               5.2 &            11.0 &    3 &   $a.$ &            $h.$ &         \nodata \\
 NGC4536 &  16.2 &      66.0 &     305.6 &                 16.7 &               4.4 &            10.2 &    3 &   $c.$ &            $k.$ &         \nodata \\
 NGC4559 &   8.9 &      65.0 &     328.0 &                 13.7 &               3.5 &             9.8 &    5 &   $g.$ &            $o.$ &         \nodata \\
 NGC4569 &  15.8 &      70.0 &      18.0 &                 20.9 &               5.9 &            10.8 &    2 &   $c.$ &            $k.$ &         \nodata \\
 NGC4625 &  11.8 &      47.0 &     330.0 &                  2.4 &               1.2 &             9.1 &    9 &   $d.$ &            $k.$ &            $r.$ \\
 NGC4651 &  16.8 &      50.1 &      73.8 &                  9.5 &               2.4 &            10.3 &    5 &   $f.$ &            $h.$ &            $r.$ \\
 NGC4689 &  15.0 &      38.7 &     164.1 &                  8.3 &               4.7 &            10.1 &    5 &   $c.$ &            $h.$ &         \nodata \\
 NGC4725 &  12.4 &      54.0 &      36.0 &                 17.5 &               6.0 &            10.8 &    1 &   $d.$ &            $k.$ &         \nodata \\
 NGC4736 &   4.4 &      41.0 &     296.0 &                  5.0 &               0.8 &            10.3 &    1 &   $d.$ &            $m.$ &         \nodata \\
 NGC4826 &   4.4 &      59.1 &     293.6 &                  6.7 &               1.5 &            10.2 &    1 &   $c.$ &            $m.$ &         \nodata \\
 NGC4941 &  15.0 &      53.4 &     202.2 &                  7.3 &               3.4 &            10.1 &    1 &   $c.$ &            $h.$ &         \nodata \\
 NGC5055 &   9.0 &      59.0 &     102.0 &                 15.5 &               4.2 &            10.7 &    4 &   $d.$ &            $m.$ &         \nodata \\
 NGC5248 &  14.9 &      47.4 &     109.2 &                  8.8 &               3.2 &            10.3 &    3 &   $c.$ &            $n.$ &         \nodata \\
 NGC5457 &   6.6 &      18.0 &      39.0 &                 23.2 &              13.4 &            10.3 &    5 &   $d.$ &            $m.$ &            $r.$ \\
 NGC6946 &   7.3 &      33.0 &     243.0 &                 12.2 &               4.5 &            10.5 &    5 &   $d.$ &            $m.$ &            $r.$ \\
 NGC7331 &  14.7 &      76.0 &     168.0 &                 19.8 &               3.7 &            11.0 &    4 &   $d.$ &            $m.$ &         \nodata \\
 NGC7793 &   3.6 &      50.0 &     290.0 &                  5.5 &               1.9 &             9.3 &    6 &   $c.$ &            $m.$ &            $r.$ \\
\bottomrule
\end{tabular}
\begin{tablenotes}
    \small
    \item {\bf Notes:} (2) Distance \citep[from EDD][]{dist_galbase_EDD_TULLY09}; 
(3-4) inclination angle and position angle  \citep{1999ApJ...523..136S,DEBLOK08,LEROY09,MUNOZMATEOS09,2009ApJ...702..277M,2010A&A...511A..89C,2013ApJ...774..126M,2014ApJ...789...81F,MAKAROV14,KOCH18,LANGMEIDT_2020ApJ...897..122L}; 
(5) isophotal radius \citep{MAKAROV14}; (6) effective radius \citep{Leroy21_PHANGS-ALMA_CO}; (7) logarithmic global stellar mass \citep{LEROY19}; (8) numerical Hubble stage T; 
(9) References of CO observations.
$a.$ \citet{Kuno07_NRO_CO};
$b.$ \citet{NIETEN06};
$c.$ PHANGS-ALMA \citep{Leroy21_PHANGS-ALMA_CO};
$d.$ HERACLES \citep{LEROY09};
$e.$ \citet{GRATIER10} and \citet{DRUARD14};
$f.$ \citet{Leroy21_CO_Line_Ratios};
$g.$ COMING \citep{Sorai19_COMING};
(10) References of \HI observations.
$h.$ EveryTHINGS (P.I. Sandstrom; Chiang et al. in preparation);
$i.$ \citet{BRAUN09};
$j.$ \citet{Puche91};
$k.$ \citet{SCHRUBA11};
$l.$ \citet{KOCH18};
$m.$ THINGS \citep{WALTER08};
$n.$ PHANGS-VLA (P.I. Utomo; Sardone et al. in preparation);
$o.$ HALOGAS \citep{Heald11_HALOGAS};
(11) References of \metal measurement.
$p.$ private communication with K. Kreckel \citep[see][]{Chiang21};
$q.$ \metal radial gradients from PHANGS-MUSE \citep{Santoro22};
$r.$ lines from \citet{Zurita21} compilation.
\end{tablenotes}
\end{table*}

\subsection{Data Sets}\label{sec:data:data sets}
In this section, we describe the data sets adopted in this work. All these data sets are processed with the method later described in Section~\ref{sec:method:data processing}.\\

\noindent\textbf{\textit{Herschel} FIR.} We use \textit{Herschel} FIR data for the dust SED fitting and calculating the colour temperature of dust emission. We adopt the \textit{Herschel} FIR maps from the J. Chastenet et al. (in preparation) background-subtracted data products.
J. Chastenet et al. (in preparation) uniformly reduced the six \textit{Herschel} photometric bands for a large sample ($\sim 800$) of nearby galaxies, using the \texttt{Scanamorphos} \citep{ROUSSEL13} routine.
This includes the 70, 100, and 160~\micron\ bands from the Photoconductor Array Camera and Spectrometer \citep[PACS;][]{POGLITSCH10}, and the 250, 350, and 500~\micron\ bands from the Spectral and Photometric Imaging Receiver \citep[SPIRE;][]{GRIFFIN10}. In this work, we adopt the 70 to 250~\micron\ data products, which yields the raw {point spread function (PSF) with} FWHM $\sim18\arcsec$. We convolve the maps to a circular Gaussian PSF with FWHM of 21\arcsec, which is the `moderate Gaussian' suggested by \citet{ANIANO11}.\\

\noindent\textbf{\textit{GALEX} UV.} We use the two bands of \textit{GALEX} at $\lambda\sim154~\rm nm$ and $\lambda\sim231~\rm nm$ (hereafter FUV and NUV, respectively) to trace \Sigmasfr. We use the data products at 15\arcsec resolution from the $z$0MGS catalog \citep{LEROY19}.\\

\noindent\textbf{\textit{WISE} near- and mid-IR.} We use the data at $\lambda\sim3.4, 4.6, 12$ and 22\,\micron\ observed by \textit{WISE} to trace dust emission in the near- and mid-IR. We also use the 3.4 and 22\,\micron\ bands (hereafter W1 and W4, respectively) to trace \Sigmastar and \Sigmasfr (the latter is supplemented by the \textit{GALEX} UV data). We use the background-subtracted data products at $15\arcsec$ resolution from the $z$0MGS collaboration \citep{LEROY19}.\\

\noindent\textbf{CO rotational lines.} We use the CO rotational lines as a tracer for molecular gas. 
For the majority of our sample galaxies, 
we use the compilation of CO mapping assembled by \citet{Leroy21_PHANGS_pipeline,Leroy21_CO_Line_Ratios} from publicly available CO $J=1\to 0$ and $J=2\to 1$ data:
\begin{itemize}
    \item CO $J=1\to 0$ data from the CO Multi-line Imaging of Nearby Galaxies (COMING) survey \citep{Sorai19_COMING,Kuno07_NRO_CO}.
    \item CO $J=2\to 1$ data from HERACLES \citep{LEROY09}, the PHANGS-ALMA survey \citep{Leroy21_PHANGS-ALMA_CO}, the IRAM M33 CO(2--1) survey \citep{GRATIER10,DRUARD14}, and a new set of data observed by the IRAM 30~m telescope focused on the Virgo Cluster \citep[P.I. Schruba; processed in][]{Leroy21_CO_Line_Ratios}.
\end{itemize}
For NGC 224 (M31), we adopt the CO $J=1\to 0$ data observed with the IRAM 30-m telescope by \citet{NIETEN06}. We obtain the map from the supplementary material\footnote{\url{https://www.astro.princeton.edu/~draine/m31dust/m31dust.html}} of \citet{DRAINE14}.
All the measurements adopted in this work target the $^{12}$C$^{16}$O isotope only. The detailed reference for each target galaxy is listed in Table~\ref{tab:samples}.

For uniformity, we convert the CO $J=2\to 1$ intensity to CO $J=1\to 0$ with the mean values of the line ratio ($R_{21}$) measured for this sample in \citet{Leroy21_CO_Line_Ratios}:
\begin{equation}\label{eq:CO_R21}
I_{\mathrm{CO}\,J=1\to0} = I_{\mathrm{CO}\,J=2\to1}/R_{21},~R_{21}=0.65^{+0.18}_{-0.14}.
\end{equation}
For simplicity, $I_\mathrm{CO}$ stands for $I_{\mathrm{CO}\,J=1\to0}$ hereafter. In the main analysis, we will use the constant $R_{21}=0.65$ as the fiducial value. Meanwhile, recent studies have attempted to formulate the variation of $R_{21}$ as a function of local physical conditions, e.g. \Sigmasfr. This variation has a minor effect on this work, which will be discussed in Appendix~\ref{app:other_aCO}. \\

\noindent\textbf{\HI 21\,cm line.} We use \HI 21\,cm line emission to trace the atomic gas surface density in the ISM. Detailed references for each galaxy are listed in Table~\ref{tab:samples}. We include several new and publicly available \HI moment 0 maps. The two new surveys, EveryTHINGS and PHANGS-VLA, used the Karl G. Jansky Very Large Array (VLA).\footnote{The VLA is operated by the National Radio Astronomy Observatory (NRAO), which is a facility of the National Science Foundation operated under cooperative agreement by Associated Universities, Inc.}
The EveryTHINGS survey \citep[][in preparation]{Chiang21} observed 39 nearby galaxies with the C and D configuration of the VLA. The typical beam size of these cubes is $\sim 20\arcsec$. For the descriptions of observation design, data reduction and imaging processes, we refer the readers to \citet[][Chapter 4.A; I. Chiang et al. in preparation]{Chiang21PhD}. 
The PHANGS-VLA \citep[P.I. Utomo;][A. Sardone et al. in preparation]{Sun20} survey observed nearby galaxies in the PHANGS catalog, with similar data reduction and imaging approaches as EveryTHINGS.\\

\noindent\textbf{Oxygen abundance.} We derive the oxygen abundance as a function of galactocentric distance by adopting the radial gradient of \metal obtained from  measurements in \HII regions, where oxygen atoms locked in the solid phase are negligible \citep{Peimbert10_DustInHII}; that is, the \metal measured in \HII regions represents the total \metal.
We use the \metal gradient or \metal measurements from the following sources: (1) The PHANGS-MUSE survey \citep{Emsellem22_PHANGS-MUSE}. They use strong line measurements with the \citet{PilyuginGrebel16} S-calibration\footnote{They utilize the $S_2=I_{\rm [S~II]}\lambda 6717 + \lambda 6731/I_{\rm H\beta}$, $N_2=I_{\rm [NII]}\lambda 6548+ \lambda 6584/I_{\rm H\beta}$, and $R_3=I_{\rm [OIII]}\lambda 4959+\lambda 5007/I_{\rm H\beta}$ line intensity ratios} (hereafter PG16S). We adopt the derived radial gradients from \citet{Santoro22}. (2) The optical \HII region emission line compilation in \citet{Zurita21}. We use their catalog to calculate the PG16S \metal in \HII regions and then fit the radial \metal gradient in each galaxy. We only consider galaxies that have at least 5 measurements spanning at least $0.5R_{25}$, where $R_{25}$ is the isophotal radius at the $B$-band surface brightness $\mu_\mathrm{B}=25~\mathrm{mag~arcsec^{-2}}$.

For galaxies without \metal\ measurements in \citet{Zurita21} or \citet{Santoro22}, we use the two-step strategy proposed by \citet{Sun20} to estimate their \metal. First, we use a mass-metallicity relation\footnote{Mass-metallicity relation is the scaling relation between stellar mass and gas-phase metallicity.} to predict \metal at one effective radius ($R_\mathrm{eff}$, the radius within which half of the galaxy's luminosity is contained, measured with $I_\mathrm{W1}$). Secondly, we extend the prediction with a radial gradient of $-0.1~\mathrm{dex}/R_\mathrm{eff}$ suggested by \citet{Sanchez14}. We characterize the mass-metallicity relation following \citet[][also see \citet{Sanchez13,Sanchez19}]{Moustakas11}:
\begin{equation}\label{eq:mass-metallicity}
    \metal = a + bxe^{-x},
\end{equation}
where $x=\log(M_\star/\mathrm{M}_\odot) - 11.5$, and $a$ and $b$ are free parameters. We obtain the best fit parameters $a=8.56\pm 0.02$ and $b=0.010\pm0.002$ by the fit to the relation between \metal at $R_\mathrm{eff}$ and $\log(M_\star/\mathrm{M}_\odot)$ for the galaxies in Table~\ref{tab:samples}.

\subsection{Signal mask}\label{sec:data:mask}
In the analysis, we mask out data points with low signal-to-noise ratios in key data. The selection criteria are listed below:
\begin{itemize}
    \item We keep pixels with $>1\sigma$ detections in the \textit{WISE} and \textit{Herschel} IR bands used for dust fitting. The change in the statistics of \Td is minor compared to a $3\sigma$ mask; thus we use the $1\sigma$ mask to include more data points. This also ensures the validity of the estimated colour temperature (Section~\ref{sec:method:physical}). This masking is done by J. Chastenet et al. (in preparation).
    \item We focus on regions with CO detections. We set the detection threshold to be $I_\mathrm{CO}\sim 0.4~\mathrm{K~km~s^{-1}}$, which corresponds to $\Sigmamol\sim 1.7\times 10^6~\SigmaMassUnitKpc$.
\end{itemize}
Most pixels passing the two criteria above are in the inner galaxy. The median radius of the pixel-by-pixel measurements is $\sim 0.4 R_{25}$. There are $\sim1.6$ per cent of data points with CO detection but no \HI detection. For these pixels, we set their \HI intensity to zero and keep them in the analysis. These pixels have their \Sigmamol span $\sim 2.7\times 10^6$--$11.4\times 10^6$~\SigmaMassUnitKpc, which is at least one order of magnitude above the weakest detection of \Sigmaatom in our dataset ($\sim10^5$~\SigmaMassUnitKpc). Thus these regions should still be H$_2$-dominated even if there is undetected \HI existing.

\section{Method}\label{sec:method}
\subsection{Multi-wavelength Data Processing}\label{sec:method:data processing}
All multi-wavelength data are convolved to a Gaussian PSF with an FWHM corresponding to 2\,kpc, using the \texttt{astropy.convolution} package \citep{ASTROPY13,Astropy18,Astropy22}. The images are then reprojected to a pixel size of one third of the FWHM (i.e.\ we oversample at roughly the Nyquist sampling rate) with the \texttt{astropy} affiliated package \texttt{reproject}. All the surface density and surface brightness quantities presented in this work have been corrected for inclination.

We blank a few regions where the emission from a foreground or background source may be confused with the main target. In NGC 4496A, we blank the region around NGC 4496B because while the IR maps detect the dust emission from NGC 4496B, the CO and \HI emission from NGC 4496B are not in the spectral coverage of the available cubes. For all \textit{GALEX} and \textit{WISE} maps, we blank the regions with known stars in the masks compiled in the $z$0MGS database. We interpolate the intensities in the blanked regions with a circular Gaussian kernel (FWHM = 22.5\arcsec) with the function \texttt{interpolate\_replace\_nans} in the \texttt{astropy.convolution} module. This interpolation is applied to the \textit{WISE} and \textit{GALEX} maps before convolution and reprojection.

\subsection{Physical Parameter Estimates}\label{sec:method:physical}

\noindent\textbf{Colour temperature.}
In order to directly compare our measurements with the model prediction in \citetalias{Hirashita_Chiang22}, we use the colour temperature, which is also adopted by \citetalias{Hirashita_Chiang22}, as our fiducial dust temperature in the analysis. We calculate the colour temperature at two wavelengths $\lambda_1$ and $\lambda_2$, $\Tclr(\lambda_1,~\lambda_2)$, in each pixel by solving the following equation:
\begin{equation}\label{eq:Tclr}
    \frac{\kappa(\lambda_1) B_\nu\Big(\lambda_1,~\Tclr(\lambda_1,~\lambda_2)\Big)}{\kappa(\lambda_2) B_\nu\Big(\lambda_2,~\Tclr(\lambda_1,~\lambda_2)\Big)} = \frac{I_\nu(\lambda_1)}{I_\nu(\lambda_2)},
\end{equation}
where $\kappa(\lambda)$ is the dust emissivity as a function of wavelength $\lambda$, and $I_\nu(\lambda)$ is the measured SED. We adopt a power-law approximation for $\kappa(\lambda)$ as
\begin{equation}\label{eq:kappa}
    \kappa(\lambda) = \kappa_0 \left(\frac{\lambda_0}{\lambda}\right)^\beta,
\end{equation}
where $\lambda_0$ is the reference wavelength, $\kappa_0$ is the emissivity at $\lambda_0$ and $\beta$ is the power-law index. This approximation is valid in the FIR and used in MBB models \citep[e.g.][]{Schwartz1982,HILDEBRAND83_MBB,GORDON14,CHIANG18,UTOMO19}. We adopt $\beta=2$, $\lambda_0=160~\micron$ and $\kappa_0=10.1~\mathrm{cm^2~g^{-1}}$ \citep{CHIANG18}. Since the emissivity described in equation~(\ref{eq:kappa}) is only used in calculating \Tclr (equation~\ref{eq:Tclr}), the values of $\lambda_0$ and $\kappa_0$ do not affect the results in this work.

Our methodology for calculating $\Tclr(\lambda_1,~\lambda_2)$ is the same as the one used in \citetalias{Hirashita_Chiang22}, except that we select a set of $(\lambda_1,~\lambda_2)$ more suitable for nearby galaxy observations. Specifically, we adopt $\lambda_1=100~\micron$ and $\lambda_2=250~\micron$ from \textit{Herschel} PACS 100~\micron\ and SPIRE 250~\micron\ data.
We adopt the same set of wavelengths when we calculate $\Tclr$ using the \citetalias{Hirashita_Chiang22} model, which is to be compared with the observationally derived $\Tclr$.
As shown in \citetalias{Hirashita_Chiang22}, the colour temperature is not sensitive to the choice of wavelengths as long as we choose two wavelengths in the FIR range, which is appropriate for the \textit{Herschel} bands. The \Tclr scales with dust temperature derived from dust SED fitting, which we will discuss in Appendix~\ref{app:Tdust-Tracers}.\\

\noindent\textbf{Dust surface density and ISRF.} 
We adopt the maps of dust properties from J. Chastenet et al. (in preparation). J. Chastenet et al. (in preparation) fit the \textit{WISE} (3.4, 4.6, 12 and 22\,\micron) and \textit{Herschel} (70, 100, 160 and 250\,\micron) dust emission SED with the \citet{DRAINE07} physical dust model \citep[with the renormalized dust opacity derived in][]{Chastenet21_M101}. The fitting is done with the Bayesian fitting tool \texttt{DustBFF} \citep{GORDON14}. The data products include maps of dust mass surface density (\Sigmad), ISRF, and the fraction of polycyclic aromatic hydrocarbons ($q_\mathrm{PAH}$). We use \Sigmad in our main analysis and the ISRF in Appendix~\ref{app:Tdust-Tracers}. Like the \textit{Herschel} maps, the dust maps have raw resolution FWHM $\sim18\arcsec$, and we convolve them to a circular Gaussian PSF with FWHM of 21\arcsec.\\

\noindent\textbf{SFR surface density.} 
We use UV+IR `hybrid' tracers for \Sigmasfr with coefficients calibrated with GSWLC data \citep{Salim16_GSWLC,Salim18_GSWLC} in the $z$0MGS project \citep{LEROY19}. This methodology corrects the dust attenuation in the UV with IR data, thus it offers better constraint on SFR than single-band tracers. We use \textit{GALEX} FUV or NUV and \textit{WISE} W4 data to trace the \Sigmasfr. For galaxies with both FUV and W4 available, we use:
\begin{multline}
\frac{\Sigmasfr}{1~\SigmasfrUnit} = \\
8.85 \times 10^{-2} \frac{I_{\rm FUV}}{1~\SEDUnit} + 3.02 \times 10^{-3} \frac{I_{\rm W4}}{1~\SEDUnit}~.
\end{multline}

For NGC 3596, for which FUV is unavailable but NUV and W4 are available, we use
\begin{multline}
\frac{\Sigmasfr}{1~\SigmasfrUnit} = \\
8.94 \times 10^{-2} \frac{I_{\rm NUV}}{1~\SEDUnit} + 2.63 \times 10^{-3} \frac{I_{\rm W4}}{1~\SEDUnit}~.
\end{multline}

For NGC 3953 and NGC 4689, for which only W4 is available, we use
\begin{equation}
\frac{\Sigmasfr}{1~\SigmasfrUnit} =
3.81 \times 10^{-3} \frac{I_{\rm W4}}{1~\SEDUnit}~.
\end{equation}

While it is a frequently used strategy, we remind the readers that there is possible higher-order polynomial dependence on the IR term \citep[e.g.][]{Buat05}, and that the coefficients for the IR term depends on the dust attenuation data or model used for calibration.
For further discussions about the advantages and disadvantages of SFR tracers, we refer the readers to the \citet[][Chaper 3]{KENNICUTT12} review.\\

\noindent\textbf{Stellar mass surface density.} We use \textit{WISE} W1 data to trace stellar mass surface density (\Sigmastar) with the conversion formula suggested by $z$0MGS \citep[][see their Appendix A]{LEROY19}:

\begin{equation}
\frac{\Sigmastar}{1~\SigmaMassUnitKpc} = 3.3 \times 10^8 \left(\frac{\Upsilon^{3.4}_\star}{0.5~\mathrm{M_\odot~L_\odot^{-1}}}\right) \frac{I_{\rm W1}}{1~\SEDUnit}~,
\end{equation}
where $\Upsilon^{3.4}_\star$ is stellar-to-W1 mass-to-light ratio determined from the specific SFR (sSFR)-like quantity calibration (using \Sigmasfr-to-$I_\mathrm{W1}$) described in \citet{LEROY19}.\\

\noindent\textbf{Metallicity.} We use the oxygen abundance, \metal, to trace metallicity ($Z$). We assume a fixed abundance pattern. The conversion from \metal to metallicity relative to solar abundance is described as
\begin{equation}
    Z' = 10^{\metal - 8.69},
\end{equation}
where $Z'$ indicates the metallicity normalized to solar, and 8.69 is the solar value of \metal \citep{ASPLUND09}.\\

\noindent\textbf{Atomic Gas Surface Density.} We trace the atomic gas surface density (\Sigmaatom) with the \HI 21\,cm line emission ($I_\HI$) data, assuming that the opacity is negligible \citep[e.g.][]{WALTER08}: 
\begin{equation}\label{eq:hi}
    \frac{\Sigmaatom}{1~\SigmaMassUnitKpc} = 
    1.36\times(1.46\times10^{4})\frac{I_\HI}{1~\ICOUnit}~,
\end{equation}
where the factor $1.36$ accounts for the mass of helium. Note that in this work, we use the unit $\SigmaMassUnitKpc$ for surface densities for more straightforward comparison with simulations instead of $\SigmaMassUnit$, which is more commonly used in observations.\\

\noindent\textbf{Molecular Gas Surface Density.} We calculate the molecular gas surface density (\Sigmamol) with the integrated intensity of CO $J=1\to0$ line ($I_\mathrm{CO}$) and a \CO-to-\HTWO conversion factor (\aco):\footnote{The conventional MW conversion factor is $\aco = 4.35$~\acoUnit (including a factor of 1.36 for helium mass). This is equivalent to $X_\CO = 2\times 10^{20}\,\XcoUnit$ in column density units (column density for $\mathrm{H}_2$ only). \aco
can be converted to $X_\CO$ units by multiplying by a factor of 
$(\mathrm{M}_{\sun}/2\cdot 1.36m_\mathrm{H})(\mathrm{cm}^2/\mathrm{pc}^2)=4.6\times 10^{19}$, where $m_\mathrm{H}$ is the mass of hydrogen atom.}
\begin{multline}\label{eq:h2}
    \frac{\Sigmamol}{1~\SigmaMassUnitKpc} = \\
    \frac{\aco}{1~\acoUnit}\times \frac{I_\mathrm{CO}}{1~\ICOUnit}\times10^6~,
\end{multline}
where the $10^6$ factor converts $\SigmaMassUnit$ to $\SigmaMassUnitKpc$. Throughout the paper, \aco is quoted for the CO $J=1\to 0$ line at 115\,\GHz and includes a factor of 1.36 to account for helium.

We can then calculate the gas surface density (\Sigmagas) as:
\begin{equation}\label{eq:Sigmagas}
    \Sigmagas = \Sigmaatom + \Sigmamol.
\end{equation}
We also calculate the total surface density (\Sigmatot), which is the combination of surface densities of stellar mass and neutral gas mass (neglecting dust mass, following \citet{BOLATTO13}) for calculating \aco later:
\begin{equation}\label{eq:Sigmatot}
    \Sigmatot = \Sigmagas + \Sigmastar .
\end{equation}

How \aco depends on local conditions is still an active field of study. In a simplified picture, the value of \aco depends on the fraction of molecular gas without CO emission \citep[the CO-dark gas; e.g.][]{WOLFIRE10,GloverMacLow11,LEROY11} and the temperature, opacity, and density of the molecular gas. The latter terms often combine to yield enhanced CO emission in galaxy centres \citep[e.g.][]{SANDSTROM13} and (ultra)luminous IR galaxies. We refer the readers to the \citet{BOLATTO13} review for a full discussion of \aco. In \citet{Chiang21}, we have demonstrated that to obtain spatially resolved dust-to-metals ratio (and thus D/G) with reasonable dependence on the local environment, one needs to adopt a conversion factor prescription that takes into account the enhancement of CO emission in galaxy centres. In this work, we adopt the \aco formula suggested by \citet{BOLATTO13}, which includes the enhancement of CO emission with total surface density:
\begin{multline}\label{eq:acoBolatto}
    \frac{\aco}{1~\acoUnit} = \\
    \rm 2.9\times\exp\left(\frac{0.4}{Z'}\right)\times\left\{
    \begin{array}{ll}
        \left(\Sigma_{\rm Total}^{100}\right)^{-0.5} & ,\,\Sigma_{\rm Total}^{100} \geq 1 \\
        1 & ,\,\Sigma_{\rm Total}^{100} < 1
    \end{array}\right.\,,
\end{multline}
where $\Sigma_{\rm Total}^{100}$ is \Sigmatot in units of $100\,\SigmaMassUnit$. Note that we calculate \aco iteratively because \Sigmatot depends on \aco. We start the loop with an initial value of $\aco=4.35~\acoUnit$ and iterate until the current \aco converges to within 1\% of the previous one. We will examine the possible changes to our results with other \aco prescriptions in Appendix~\ref{app:other_aCO}.\\

\noindent\textbf{Dust-to-Gas Ratio.}
We calculate the D/G as
\begin{equation}
    {\rm D/G} \equiv \Sigmad/\Sigmagas.
\end{equation}

\subsection{Data Weighting}\label{sec:method:weight}
In this work, we perform pixel-by-pixel analyses with our measurements. For the key quantity in this paper, dust temperature, we weight the measurements by their \Sigmad in the calculations of all statistical quantities, i.e.\ percentiles, regressions and correlation coefficients, to have the statistics reflecting the averaged properties for the dust component. For all the other analysis, we use a uniform weighting in the statistics (mainly the scaling relation between \Sigmasfr and \Sigmad in Section~\ref{sec:measure:KS law}).

\subsection{Theoretical models for dust temperature \texorpdfstring{\citepalias{Hirashita_Chiang22}}{(Paper I)}}\label{sec:method:model}
In Section \ref{sec:comparison}, we compare our measurements to the dust temperature predicted by the model described in \citetalias{Hirashita_Chiang22}. Here we briefly review the model and refer the reader to \citetalias{Hirashita_Chiang22} for details. To derive the relations between dust temperature and local physical conditions, we consider the following two models that can be treated analytically: (i) radiative transfer (RT) and (ii) one-temperature (one-$T$) models.
In the RT model, we put stars in the midplane of the disc and the dust in a screen geometry. In this model, the temperature gradient in the direction perpendicular to the disc plane naturally emerges, so that the model is suitable for investigating a multi-temperature effect for the dust. 
In the one-$T$ model, we assume that dust and stars are well mixed, so that the dust temperature is assumed to have a single value determined by the global balance between the absorbed and radiated energy by the dust.
These two extremes serve to bracket the most realistic scenario.
The radiation field from stars is given based on the gas mass surface density assuming the Kennicutt-Schmidt law \citep[hereafter the KS law]{Kennicutt98,KENNICUTT12,Kennicutt21}. The KS law links \Sigmasfr and \Sigmagas through the following formula:
\begin{equation}\label{eq:KS law}
    \log\Bigg(\frac{\Sigmasfr}{1~\SigmasfrUnit}\Bigg) = \log A  + N\log\Bigg(\frac{\Sigmagas}{1~\SigmaMassUnitKpc}\Bigg),
\end{equation}
where $A$ and $N$ are free parameters.

In the RT model, we solve the radiative transfer equation in plane-parallel geometry under given values of \Sigmagas and D/G. For simplicity, we assume a uniform disc that extends infinitely in the plane of the galaxy (or equivalently the disc is much thinner than its horizontal extent). As mentioned in \citetalias{Hirashita_Chiang22}, this simplification does not have any essential impact on the results, since the corresponding geometric factor only weakly affects the resulting dust temperature. 
The intrinsic SED of the stellar population is calculated based on a continuous, constant \Sigmasfr throughout a period $\tau_\star$, using \textsc{starburst99} \citep{Leitherer99_Starburst99} with standard parameters. Starting with this intrinsic SED at the mid-plane of the disc, the RT model solves the radiative transfer equation for dust absorption in the direction perpendicular to the disc plane (i.e.\ in each dust layer parallel to the disc plane), and the dust temperature in each dust layer is calculated based on the radiative equilibrium. Finally, the dust emission from all layers are integrated to obtain the total dust emission SED, $\mathcal{I}_\mathrm{dust}^\mathrm{RT}(\nu)$ (surface luminosity). The colour temperature of the dust emission SED is calculated by equation (\ref{eq:Tclr}) by substituting $I_\nu(\lambda)$ with $\mathcal{I}_\mathrm{dust}^\mathrm{RT}(\nu)$.

The results of these models are briefly summarized as follows. With a given value of D/G, the dust temperature rises with \Sigmasfr and \Sigmad (which also increases with \Sigmasfr because of the KS law). The \Tclr--\Sigmasfr and \Tclr--\Sigmad relations depend on D/G in such a way that lower values of D/G lead to higher dust temperatures.
The RT and one-$T$ models predict similar colour temperatures with the same set of \Sigmad and \Sigmasfr. In particular, the results of the two models are almost identical in the parameter ranges appropriate for the sample in this paper. Thus, we concentrate on the RT model only, and we provide a summary of the model setup in what follows.

In this work, there are some modifications from \citetalias{Hirashita_Chiang22}. Considering that the typical stellar ages of our nearby sample is older than those of the high-redshift sample in \citetalias{Hirashita_Chiang22}, we recalculate the intrinsic stellar SED with $\tau_\star =10$ Gyr, which is a typical age of nearby galaxies. Indeed, this choice of $\tau_\star$ is equivalent to $\mathrm{sSFR}\sim 10^{-10}$ yr$^{-1}$, which is consistent with our sample (Section~\ref{sec:comparison:implication}).
However, as shown in \citetalias{Hirashita_Chiang22}, the resulting dust temperatures are not sensitive to $\tau_\star$, so that the change of $\tau_\star$ by a factor of 2 (comparable to the dispersion of sSFR) does not affect our discussions and conclusions.

Another modification from \citetalias{Hirashita_Chiang22} is the adopted KS law coefficients.
With $\Sigmad=\mathrm{(D/G)}\Sigmagas$ and the KS law, we connect the two key quantities that govern the dust temperature in the model, \Sigmasfr and \Sigmad, as $\Sigmasfr\propto[\Sigmad/\mathrm{(D/G)}]^N$. For the consistency with the adopted data set in this work, instead of using the fiducial coefficients in \citetalias{Hirashita_Chiang22},\footnote{\citetalias{Hirashita_Chiang22} adopted $A=1.0\pm0.3\times 10^{-12}$ and $N=1.4$.} we will use the coefficients and functional form that fit best for our data set. The details will be described in Section~\ref{sec:measure:KS law}.

As noted above, we change one of the wavelengths selected for calculating the colour temperature (Section~\ref{sec:method:physical}). In \citetalias{Hirashita_Chiang22}, we used $\lambda_1=100~\micron$ and $\lambda_2=200~\micron$ as the wavelengths are near to ALMA bands for galaxies at $z>5$. In this work, we select wavelengths $\lambda_1=100~\micron$ and $\lambda_2=250~\micron$ for the comparision with the photometry data in \textit{Herschel} PACS and SPIRE bands, respectively.

\section{Results}\label{sec:measurements}

\subsection{The measured dust temperature}\label{sec:measurements:Tclr-SigmaSFR}

We present the measurements for dust temperature (traced with \Tclr) in this subsection. We first investigate the primary dependence of \Tclr on the local physical conditions, and then look for possible secondary dependence once the primary dependence is removed.

\begin{table}
\centering
\caption{Correlation between \Tclr and local conditions. We calculate the Pearson's correlation coefficient ($\rho$) along with the $p$-value of the correlation.}
\label{tab:Td_corr}
\begin{tabular}{ccc}
\toprule
                                          Quantity &  Pearson's $\rho$ with \Tclr & $p$-value \\
\midrule
           \Sigmaatom & 0.13 & $\ll 0.001$ \\
 $\log(\Sigmamol)$ & 0.67 & $\ll 0.001$ \\
 $\log(\Sigmagas)$ & 0.66 & $\ll 0.001$ \\
 $\log(\Sigmasfr)$ & 0.89 & $\ll 0.001$ \\
 $\log(\Sigmad)$ & 0.61 & $\ll 0.001$ \\
 $\log(\mathrm{D/G})$ & 0.05 & 0.543 \\
 $\log(\Sigmastar)$ & 0.67 & $\ll 0.001$ \\
 \metal & 0.49 & $\ll 0.001$ \\
$R_\mathrm{g}/R_{25}$ & $-0.58$ & $\ll 0.001$ \\
\bottomrule
\end{tabular}
\begin{tablenotes}
    \small
    \item {\bf Notes:} (a) We consider the correlation as significant when we have $p<0.05$.
\end{tablenotes}
\end{table}

\begin{figure}
	\includegraphics[width=\columnwidth]{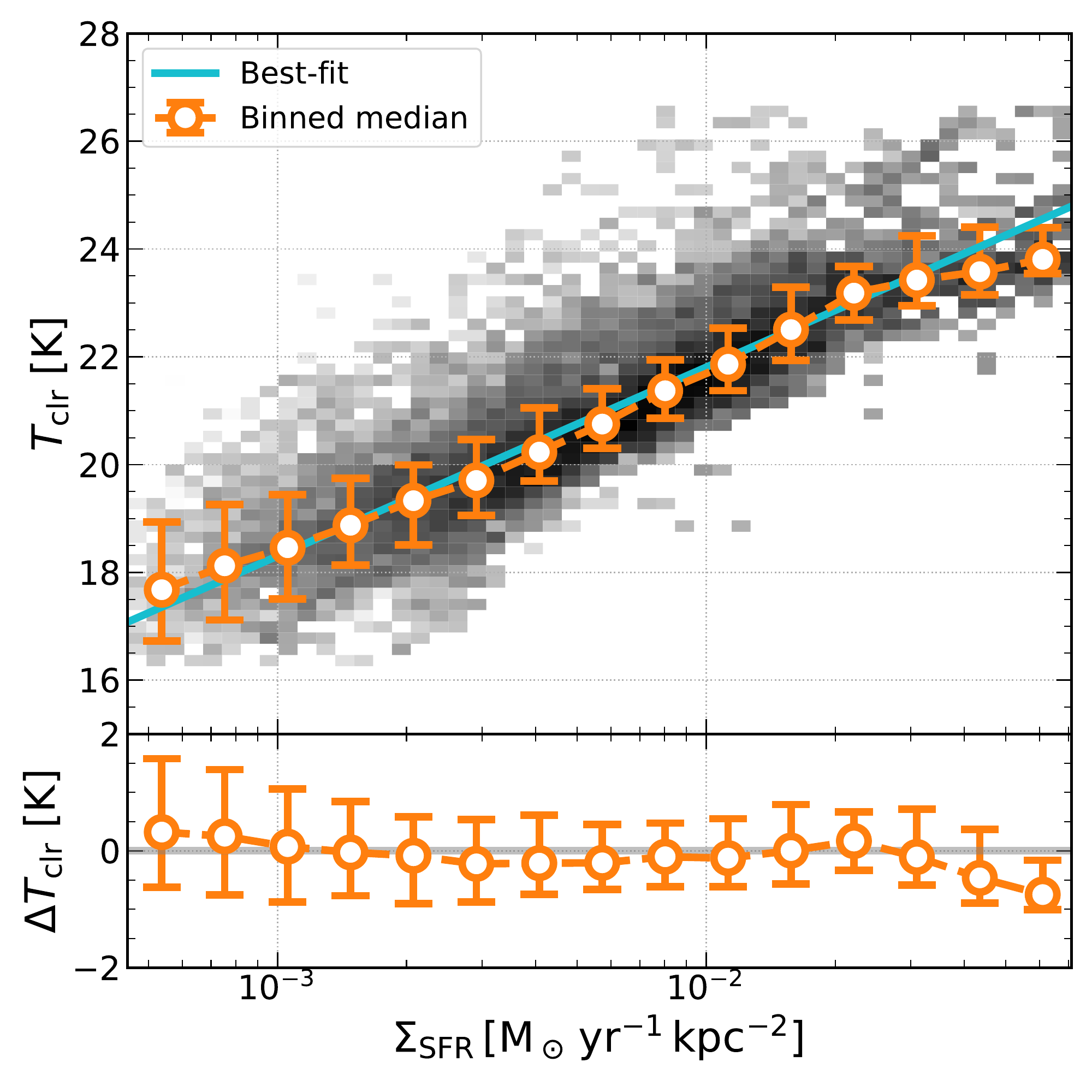}
    \caption{Measured scaling relation between \Tclr and \Sigmasfr. Top panel: \Tclr displayed as a function of \Sigmasfr. The gray scale shows the density of data points. The orange circles show the median \Tclr in each \Sigmasfr bin, and the errorbars show the 16th- and 84th-percentiles. The cyan line shows the best-fit power law between \Tclr and \Sigmasfr (equation \ref{eq:Tclr-SigmaSFR}).
    Bottom panel: $\Delta$\Tclr (defined in equation~\ref{eq:dT}) displayed as a function of \Sigmasfr, with a gray line showing zero position.}
    \label{fig:2kpc_SigmaSFR_T(250_100)_new}
\end{figure}

In Table~\ref{tab:Td_corr}, we show Pearson's correlation coefficient ($\rho$) between \Tclr and each of the quantities available in our analysis. Since most quantities have certain radial dependence, we include $R_\mathrm{g}/R_{25}$ as a reference quantity. Quantities that have correlations with \Tclr stronger than $R_\mathrm{g}/R_{25}$ are more likely to have first-order correlations with \Tclr.
Among the quantities with correlation stronger than $R_\mathrm{g}/R_{25}$, \Sigmasfr has the strongest correlation with \Tclr ($\rho=0.89$, showing a strong correlation). This agrees with the baseline assumption in \citetalias{Hirashita_Chiang22} that the UV radiation from young stars is the dominant dust heating source in the majority of the sample. \Sigmamol, \Sigmagas, \Sigmad, and \Sigmastar also have correlations with \Tclr stronger than $R_\mathrm{g}/R_{25}$. For \Sigmamol and \Sigmagas, both correlations likely result from their strong correlations with \Sigmasfr, i.e the KS law. The correlation between \Sigmastar and \Tclr also likely result from the correlation between \Sigmastar and \Sigmasfr, possibly indicating the resolved star-forming main sequence \citep{Cano-Diaz16,Abdurro'uf17,Abdurro'uf18,Hsieh17,Liu18,Maragkoudakis17,Medling18,Lin19,Lin22,Morselli20,Ellison21_ALMaQUEST_V,Pessa22}.
\Sigmad has a correlation slightly stronger than the radial dependence, which might result from its correlation with \Sigmagas due to the relatively small D/G variations. Note that from the analysis in \citetalias{Hirashita_Chiang22}, we expect a negative correlation between \Sigmad and \Tclr at fixed \Sigmasfr because of the effect of dust shielding. We conversely observe a moderately positive correlation between \Sigmad and \Tclr here, since with increasing \Sigmad, \Sigmasfr (or the UV radiation field) increases owing to the KS law.

We display \Tclr as a function of \Sigmasfr in the top panel of Fig.~\ref{fig:2kpc_SigmaSFR_T(250_100)_new}. As shown in the figure and implied from the correlation, the observed \Tclr seems to have a power-law dependence on \Sigmasfr. The best-fit power-law relation is
\begin{equation}\label{eq:Tclr-SigmaSFR}
    \frac{\Tclr}{1~\mathrm{K}} = (3.50\pm0.02)\log\Bigg(\frac{\Sigmasfr}{1~\SigmasfrUnit}\Bigg) + (28.81\pm0.04)~.
\end{equation}
The root-mean-square deviation (RMSD) between measured \Tclr and the best-fit relation is 0.81~K.\footnote{If we only include pixels with IR measurements $>3\sigma$ instead of $1\sigma$ (Section~\ref{sec:data:mask}), the coefficient, offset and RMSD in equation~\ref{eq:Tclr-SigmaSFR} change to $3.43\pm0.02$, $28.68\pm0.05$ and 0.80, respectively. These changes are minor compared to the error bars in Fig.~\ref{fig:2kpc_SigmaSFR_T(250_100)_new}.} The residual between the measurements and best-fit relation $\Delta\Tclr$ is defined as 
\begin{equation}\label{eq:dT}
    \Delta\Tclr \equiv T_\mathrm{clr}^\mathrm{measured} - T_\mathrm{clr}^\mathrm{best\mbox{-}fit}(\Sigmasfr),
\end{equation}
where the first and second terms on the right-hand side are the measured and best-fit colour temperatures, respectively. We show $\Delta$\Tclr as a function of \Sigmasfr in the bottom panel of Fig.~\ref{fig:2kpc_SigmaSFR_T(250_100)_new}. The residual $\Delta$\Tclr is centred around zero throughout almost the entire observed \Sigmasfr range, again indicating the significance of the power-law relation between \Tclr and \Sigmasfr. However, we do notice that there tends to be a larger scatter toward the low-\Sigmasfr end, and larger deviation ($>0.5$~K) toward the high-\Sigmasfr end.

\begin{table}
\centering
\caption{Correlation between $\Delta$\Tclr and local conditions. We calculate Pearson's correlation coefficient ($\rho$) along with the $p$-value of the correlation.}
\label{tab:dT_corr}
\begin{tabular}{lcc}
\toprule
Quantity & Pearson's $\rho$ with $\Delta$\Tclr & $p$-value \\
 \midrule
 \Sigmaatom & 0.21 & $\ll 0.001$ \\
 $\log(\Sigmamol)$ & -0.12 & $\ll 0.001$ \\
 $\log(\Sigmagas)$ & -0.05 & $\ll 0.001$ \\
 $\log(\Sigmad)$ & -0.26 & $\ll 0.001$ \\
 $\log(\mathrm{D/G})$ & -0.27 & $\ll 0.001$ \\
 $\log(\Sigmastar)$ & 0.06 & $\ll 0.001$ \\
 \metal & -0.01 & 0.012 \\
 $R_\mathrm{g}/R_{25}$ & -0.1 & $\ll 0.001$\\
\bottomrule
\end{tabular}
\begin{tablenotes}
    \small
    \item {\bf Notes:} (a) $\Delta$\Tclr is defined in equation~(\ref{eq:dT}). (b) \Sigmasfr is excluded from this table because it has no correlation with $\Delta$\Tclr, as expected. (c) We consider the correlation as significant when we have $p<0.05$.
\end{tablenotes}

\end{table}

To investigate the possible secondary dependence of \Tclr on the local environment, we calculate the correlations between $\Delta$\Tclr and local conditions, and present them in Table~\ref{tab:dT_corr}. All the quantities show weak or no correlations with $\Delta$\Tclr, indicating that there is no single secondary parameters that strongly affects \Tclr throughout all observed samples.

\begin{figure}
	\includegraphics[width=\columnwidth]{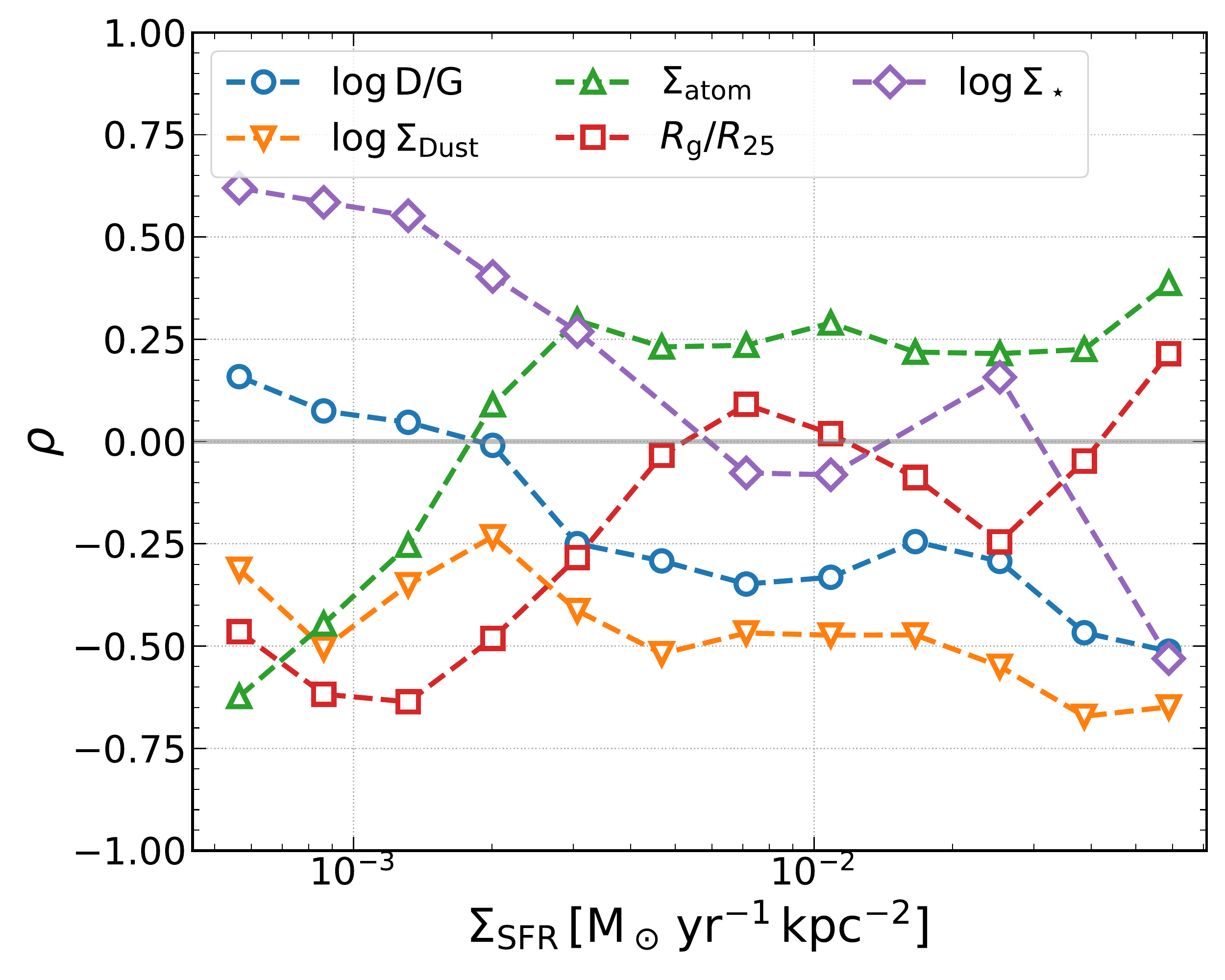}
    \caption{Pearson's correlation coefficient ($\rho$) between $\Delta$\Tclr (equation~\ref{eq:dT}) and selected local environment quantities in each \Sigmasfr bin. Quantities made into this selection have their $|\rho|$ values exceeding 0.5 in at least one of the bins. We skip the bins with $p$-values greater than 0.05, which happens in a few high-\Sigmasfr bins for $\log\Sigmastar$.}
    \label{fig:dT-corr}
\end{figure}

To further approach the possible secondary dependence, we group the data points in bins of \Sigmasfr, and calculate the correlation between $\Delta$\Tclr and local conditions in each bin. The result is shown in Fig.~\ref{fig:dT-corr}, only keeping the quantities that have their $|\rho|>0.5$ in at least one of the \Sigmasfr bins.
D/G and \Sigmad, which are expected to reduce \Tclr due to increased opacity from the \citetalias{Hirashita_Chiang22} analysis, show stronger negative correlations with $\Delta$\Tclr toward higher \Sigmasfr. \Sigmad has a stronger correlation than D/G and the correlation coefficient is negative throughout the observed range, suggesting that \Sigmad traces the dust opacity effect on \Tclr more directly in observations than D/G. On the other hand, D/G has a weak but positive correlation coefficient at low \Sigmasfr, which means that there are other mechanisms that cancel out the opacity effect traced by D/G.

The correlation between $\Delta$\Tclr and normalized radius is almost a mirror image of the one between $\Delta$\Tclr and $\log\Sigmastar$, reflecting the fact that the radius normalized by $R_{25}$ correlates with $\log\Sigmastar$, i.e. an exponential disc, in our sample.
Similar to the case of D/G mentioned above, the correlation between \Sigmastar and $\Delta$\Tclr behaves differently at low and high \Sigmasfr.
Below $\Sigmasfr\sim 2\times10^{-3}~\SigmasfrUnit$, \Sigmastar positively correlates with $\Delta$\Tclr. This positive correlation for \Sigmastar might indicate the dust heating contributed by old stellar population in the low-\Sigmasfr region. At mid-\Sigmasfr values, \Sigmastar weakly correlates with $\Delta$\Tclr. The correlation between \Sigmastar and $\Delta$\Tclr jumps to moderately negative strength only at the highest-\Sigmasfr bin, where the cause is unclear and is to be identified with future data focusing on high \Sigmasfr regions.
The reason for the correlation between $\Delta$\Tclr and \Sigmaatom is even more difficult to identify because it depends on various complicated factors such as the transition from atomic to molecular phases.  \Sigmaatom has a moderately negative correlation with $\Delta$\Tclr at low \Sigmasfr, which quickly weakens as \Sigmasfr increases. At mid- to high-\Sigmasfr values, \Sigmaatom has a weak positive correlation with $\Delta$\Tclr. 

\subsection{Scaling relation between the surface densities of SFR, gas mass, and dust mass}\label{sec:measure:KS law}

\begin{figure}
	\includegraphics[width=\columnwidth]{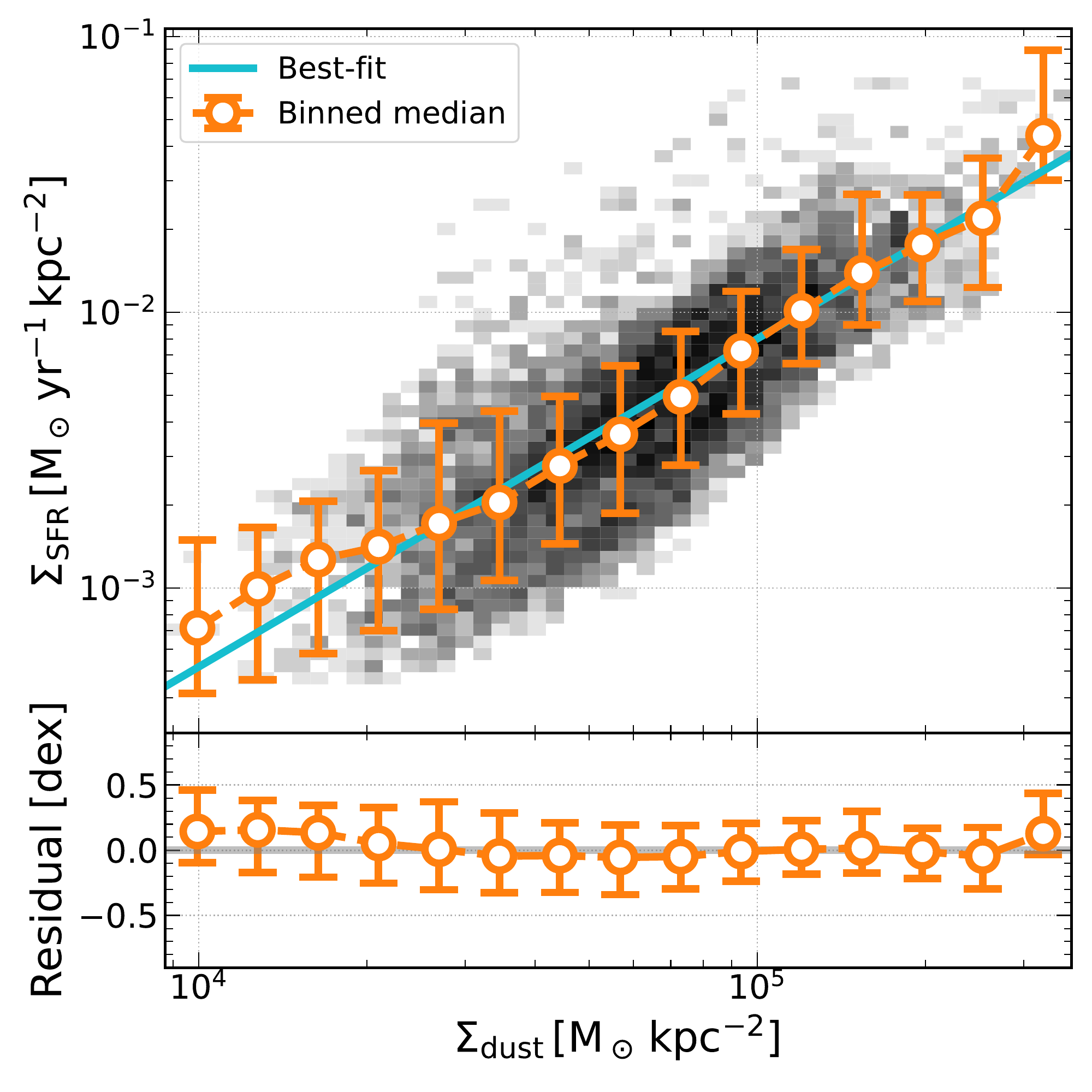}
    \caption{Measured scaling relation between \Sigmasfr and \Sigmad. Top panel: \Sigmasfr displayed as a function of \Sigmad. The gray scale shows the density of data points. The orange circles show the median \Sigmasfr in each \Sigmad bin, and the error bars show the 16th- and 84th-percentiles. The cyan line shows the best-fit power law between \Sigmasfr and \Sigmad.
    Bottom panel: The residual,  $\log(\Sigma_\mathrm{SFR}^\mathrm{measured})-\log(\Sigma_\mathrm{SFR}^\mathrm{best\mbox{-}fit})$ displayed as a function of \Sigmad.}
    \label{fig:2kpc_SigmaSFR_SigmaDust}
\end{figure}

\begin{figure*}
	\includegraphics[width=0.99\textwidth]{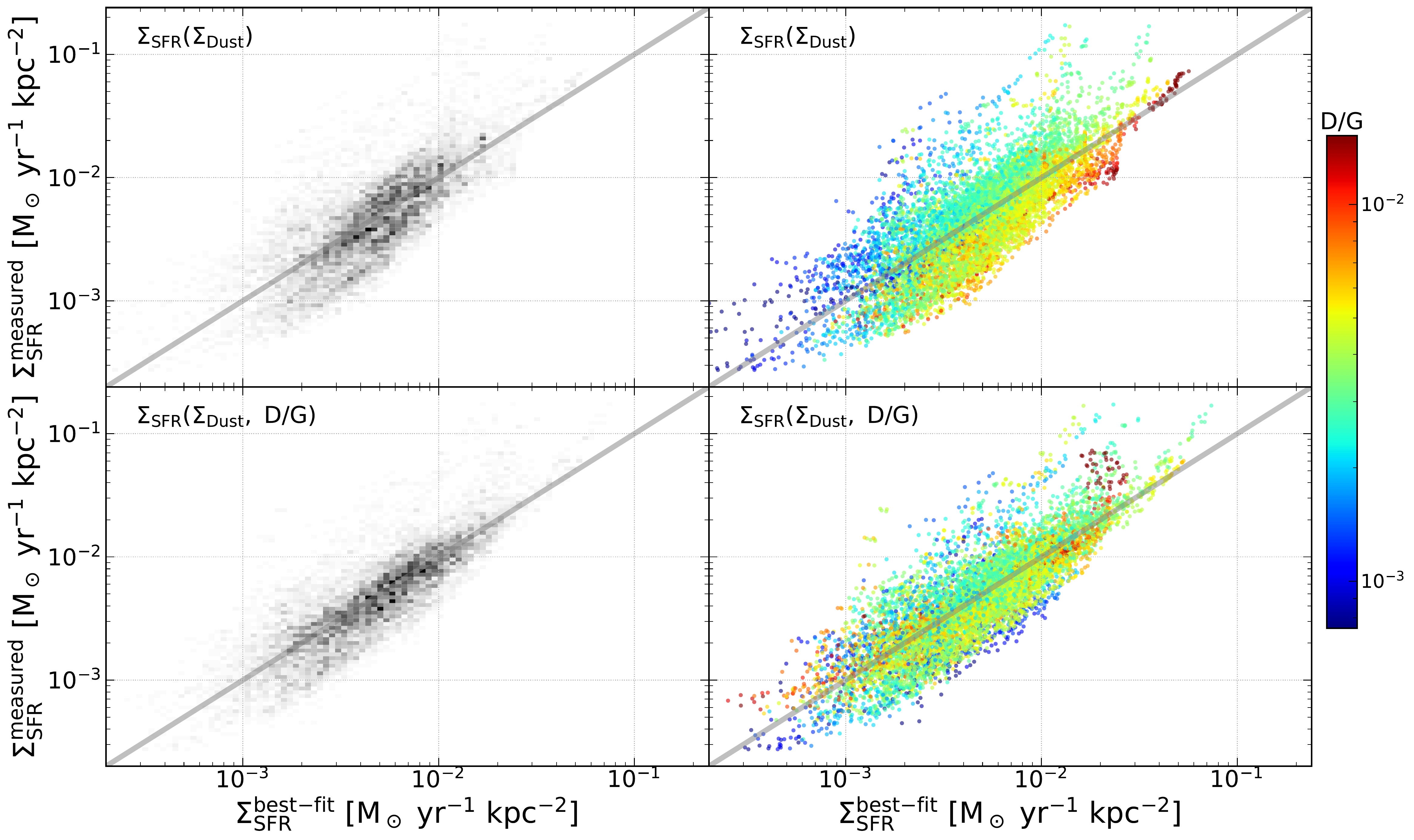}
    \caption{Comparison between measured \Sigmasfr ($\Sigma_\mathrm{SFR}^\mathrm{measured}$) and \Sigmasfr predicted from best-fit relations ($\Sigma_\mathrm{SFR}^\mathrm{best\mbox{-}fit}$) with \Sigmad alone (top panels) or with both \Sigmad and D/G (bottom panels). The left panels have their transparency indicating the distribution of data points, while the right panel shows the pixel-by-pixel data points colorized by D/G.}
    \label{fig:SigmaSFR_colorized}
\end{figure*}

As mentioned in Section~\ref{sec:method:model}, an underlying assumption in the \citetalias{Hirashita_Chiang22} model is that at a given D/G, one can infer $\Sigma_\mathrm{SFR}$ from $\Sigma_\mathrm{dust}$ through the KS law (and \Sigmagas). Thus, in this subsection, we first present the measured relation between \Sigmasfr and \Sigmad in Fig.~\ref{fig:2kpc_SigmaSFR_SigmaDust}. As shown in the figure, \Sigmasfr is correlated with \Sigmad. To build a reference for further comparison, we fit a power-law relation between the two. The best-fit relation is:
\begin{multline}
    \log\Bigg(\frac{\Sigmasfr}{1~\SigmasfrUnit}\Bigg) = \\ (1.19\pm0.01)\log\Bigg(\frac{\Sigmad}{1~\SigmaMassUnitKpc}\Bigg) - (8.05\pm0.05).
\end{multline}
The RMSD of this best-fit relation is 0.27~dex. The residual, $\log(\Sigma_\mathrm{SFR}^\mathrm{measured})-\log(\Sigma_\mathrm{SFR}^\mathrm{best\mbox{-}fit})$, has Pearson's correlation coefficient of $-0.36$ with D/G, suggesting a medium trend that \Sigmasfr increases toward regions with lower D/G at a given \Sigmad. We compare the measured and best-fit \Sigmasfr in the top-left panel of Fig.~\ref{fig:SigmaSFR_colorized}. Although we are able to get a solution minimizing the overall RMSD, the solution tends to underestimate \Sigmasfr at high \Sigmasfr and overestimate \Sigmasfr at low \Sigmasfr. In other words, the $\log(\Sigma_\mathrm{SFR}^\mathrm{best\mbox{-}fit})$ has a smaller dynamical range than $\log(\Sigma_\mathrm{SFR}^\mathrm{measured})$.

Inspired by the design of the \citetalias{Hirashita_Chiang22} model, we investigate whether we can improve the prediction of \Sigmasfr by including D/G as the second variable. This idea is supported by the top-right panel of Fig.~\ref{fig:SigmaSFR_colorized}, where we observe a systematic variation in D/G in the $\log(\Sigma_\mathrm{SFR}^\mathrm{measured})-\log(\Sigma_\mathrm{SFR}^\mathrm{best\mbox{-}fit})$ space.
The best-fit \Sigmasfr in a two-variable (\Sigmad, D/G) fitting is
\begin{multline}\label{eq:KSDust_D/G}
    \log\Bigg(\frac{\Sigmasfr}{1~\SigmasfrUnit}\Bigg) = \\ (1.49\pm0.01)\log\Bigg(\frac{\Sigmad}{1~\SigmaMassUnitKpc}\Bigg) - (0.73\pm0.01)\log(\mathrm{D/G}) \\ - (11.26\pm0.07).
\end{multline}
The RMSD of this best-fit decreases to 0.22~dex. In the bottom panels of Fig.~\ref{fig:SigmaSFR_colorized}, we use this fitting to the best-fit \Sigmasfr ($\Sigma_\mathrm{SFR}^\mathrm{best\mbox{-}fit}$). We observe that $\Sigma_\mathrm{SFR}^\mathrm{best\mbox{-}fit}$ now traces the measurements better over the entire observed range, and the systematic trend of D/G in the $\log(\Sigma_\mathrm{SFR}^\mathrm{measured})-\log(\Sigma_\mathrm{SFR}^\mathrm{best\mbox{-}fit})$ space decreases.

With the best-fit result in equation~(\ref{eq:KSDust_D/G}), we claim that the assumption in \citetalias{Hirashita_Chiang22}, i.e. scaling of \Sigmasfr with \Sigmad at given D/G, is consistent with our observation. To interpret the equation, we can convert \Sigmad in equation~(\ref{eq:KSDust_D/G}) to \Sigmagas and D/G, i.e.
\begin{multline}\label{eq:KS_D/G}
    \log\Bigg(\frac{\Sigmasfr}{1~\SigmasfrUnit}\Bigg) \sim \\ 1.49\log\Bigg(\frac{\Sigmagas}{1~\SigmaMassUnitKpc}\Bigg) + 0.76\log(\mathrm{D/G}) - 11.26.
\end{multline}
This functional form could be interpreted as a KS law with a correction term in D/G (hereafter D/G-modified KS law), with higher SFR toward larger D/G at given \Sigmagas. The 1.49 factor corresponds to the power-law index of KS law ($N$ in equation~\ref{eq:KS law}). This is consistent with the $N\sim 1.4$ value measured in \citet{Kennicutt98,de_los_Reyes19} with total gas; meanwhile, our $N$ value is equal to or steeper than the values measured with molecular-gas-only KS law \citep[$N_\mathrm{H_2}\sim 0.8$--1.5,][]{Bigiel08,Blanc09,Rahman11,LEROY13,Kennicutt21}. The D/G correction term can cause an offset up to $\sim$0.8~dex in \Sigmasfr with the observed D/G range ($\sim10^{-3}$--$10^{-2}$), which is a significant change.

The D/G can influence the star formation activity through the formation of H$_2$ on dust surfaces \citep{Krumholz08,Krumholz09,McKeeKrumholz2010}. Such a condition may be favourable for star formation through the shielding of UV radiation by H$_2$ and dust at a given \Sigmagas. However, it is difficult to prove with our data that the D/G directly affects the star formation activity, because D/G is correlated with various other physical quantities that could affect star formation, such as metallicity, gas pressure, gravitational field, etc. We refer readers who are interested in how these physical quantities could affect SFR to the \citet{KENNICUTT12} review.

\section{Comparing to the Hirashita \& Chiang (2022) model}\label{sec:comparison}

\subsection{Dependence of \texorpdfstring{\Tclr}{dust temperature} on the SFR surface density and D/G}\label{sec:comparison:Tclr_SigmaSFR}

\begin{figure}
	\includegraphics[width=\columnwidth]{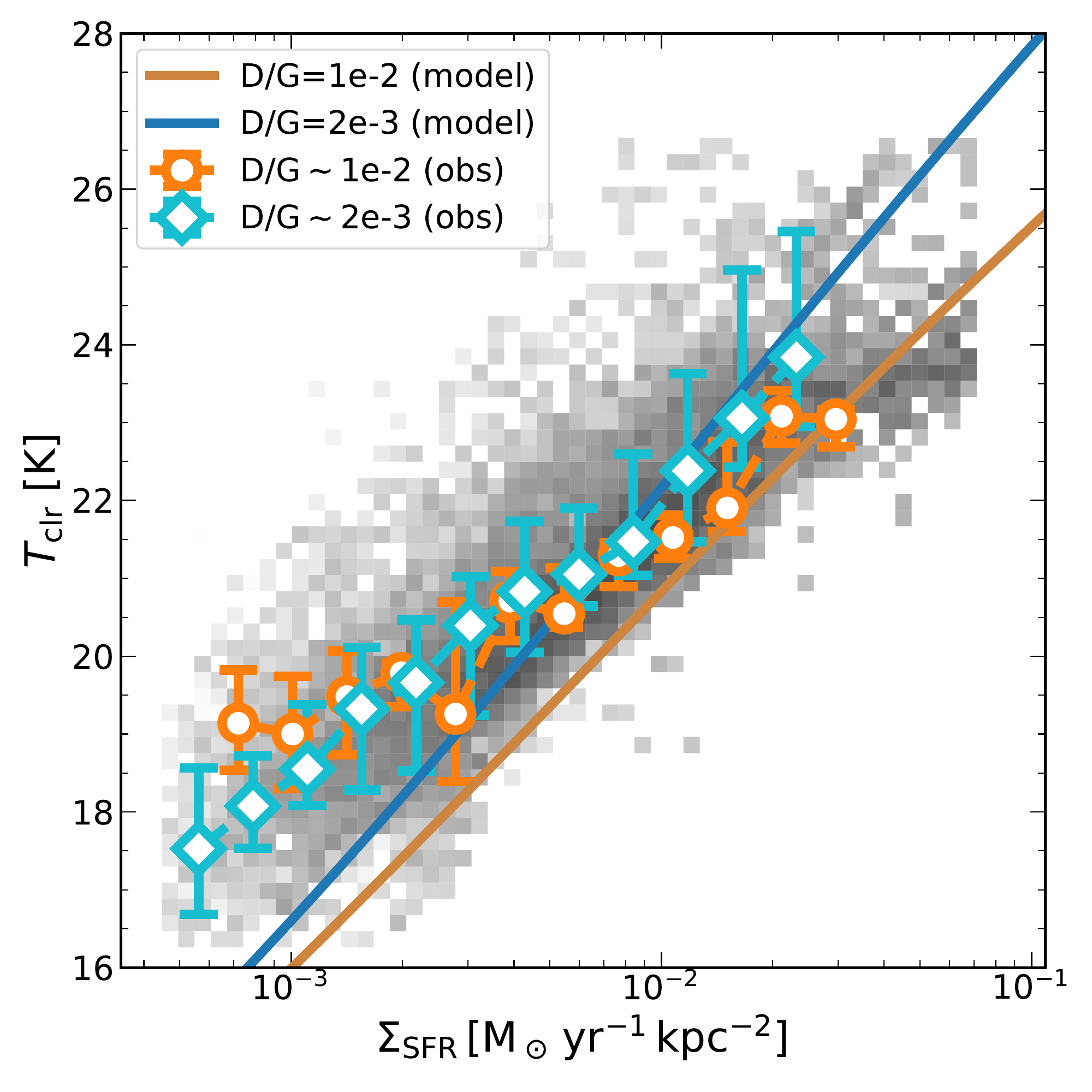}
    \caption{Comparison between the predicted and measured relations between \Tclr and \Sigmasfr. 
    The solid lines show the predicted \Tclr at two D/G values ($\mathrm{D/G}=1.0\times 10^{-2}$ and $2.0\times 10^{-3}$ for brown and blue lines, respectively) from the \citetalias{Hirashita_Chiang22} model, calculated with coefficients in equation~(\ref{eq:KS_D/G}).
    The data points with error bars present the measured \Tclr within $\pm 0.1$~dex of the desired D/G values ($1.0\times 10^{-2}$ and $2.0\times 10^{-3}$ for orange and cyan, respectively).
    The gray colour shows the distribution of all measurements (the same one shown in Fig.~\ref{fig:2kpc_SigmaSFR_T(250_100)_new}). 
    }
    \label{fig:2kpc_SigmaSFR_T(250_100)_sim}
\end{figure}

We display both measured and model-predicted \Tclr as a function of \Sigmasfr in Fig.~\ref{fig:2kpc_SigmaSFR_T(250_100)_sim}. The predictions are calculated with the \citetalias{Hirashita_Chiang22} model, the set up descriptions in Section~\ref{sec:method:model}, and the coefficients in equation~(\ref{eq:KS_D/G}).
From the \citetalias{Hirashita_Chiang22} model, we expect \Tclr to increase with \Sigmasfr, and to increase with decreasing D/G at fixed \Sigmasfr, as shown in the blue and brown lines for $\mathrm{D/G}=1.0\times 10^{-2}$ and $2.0\times 10^{-3}$, respectively. To compare with this prediction, we select data points within $\pm 0.1$~dex of the two desired D/G values and plot them with connected data points with error bars in Fig.~\ref{fig:2kpc_SigmaSFR_T(250_100)_sim}.

We observe that both the $\rm D/G\sim 1.0\times10^{-2}$ and $\rm D/G\sim 2.0\times10^{-3}$ groups follow the predicted trend at $\Sigmasfr\gtrsim 10^{-2}~\SigmasfrUnit$. Meanwhile, the separation of the two observed trends is not significant, with both groups falling within the $1\sigma$ scatter of data points of each other.
At lower \Sigmasfr, the observed \Tclr exceeds the model prediction. Moreover, the difference in \Tclr between different D/G values become even less distinguishable. This implies that some assumptions made in \citetalias{Hirashita_Chiang22} might break down at low \Sigmasfr as we will discuss in Section~\ref{sec:comparison:implication}.
Nevertheless, we emphasize that the \citetalias{Hirashita_Chiang22} model reproduces not only the overall trend but also the different relations for different values of D/G at $\Sigmasfr\gtrsim 10^{-2}~\SigmasfrUnit$.

\subsection{Dependence of \texorpdfstring{\Tclr}{dust temperature} on the surface densities of SFR and dust}\label{sec:comparison:Tclr_Tclr}

\begin{figure*}
	\includegraphics[width=\textwidth]{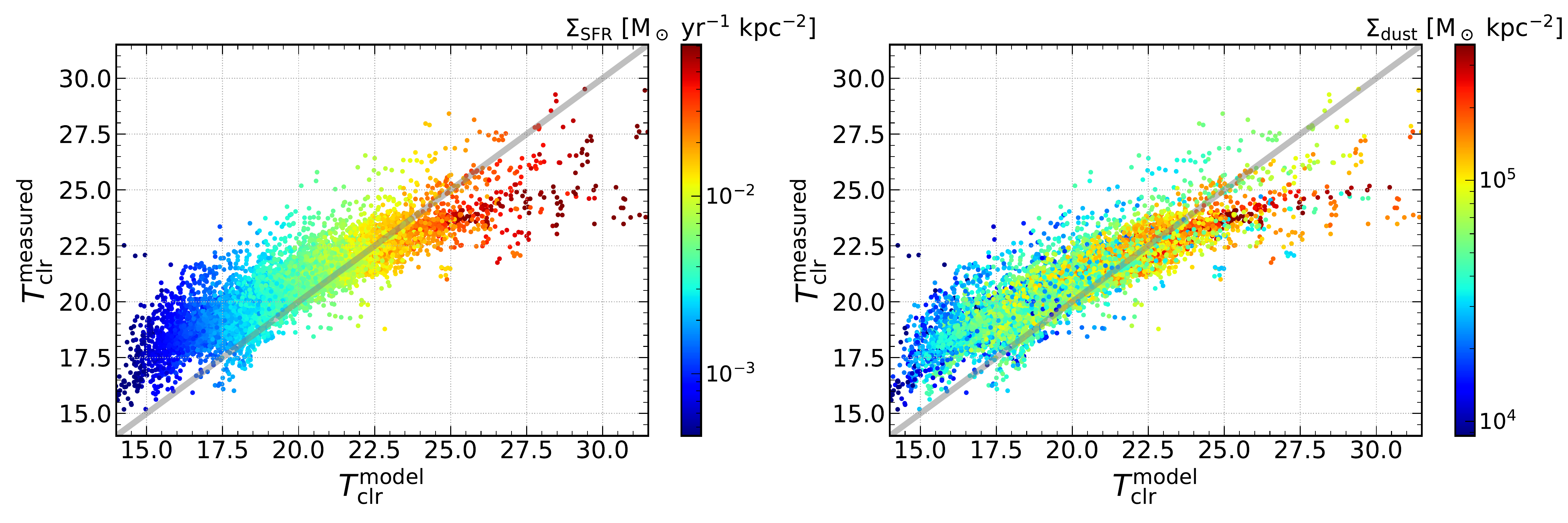}
    \caption{Measured and model-predicted \Tclr. The latter quantity, $T_\mathrm{clr}^\mathrm{model}$, is calculated with the measured pixel-by-pixel (\Sigmasfr, \Sigmad) pair using the star formation law described in equation (\ref{eq:KS_D/G}). The same data are plotted in the left and right panels, of which the data points are colour-coded by \Sigmasfr and \Sigmad, respectively.
    }
    \label{fig:Tclr-Tclr}
\end{figure*}

One interesting conclusion in \citetalias{Hirashita_Chiang22} is that with \Sigmasfr and \Sigmad, one will be able to determine dust temperature (details in their Section~3.3). To test this with our data, we calculate \Tclr from the measured \Sigmasfr and \Sigmad in this work with the \citetalias{Hirashita_Chiang22} model, and compare the model-predicted \Tclr to the measured one (the predicted data thus do not have matching \Sigmagas and D/G with observations). As shown in Fig.~\ref{fig:Tclr-Tclr}, the measured \Tclr ($T_\mathrm{clr}^\mathrm{measured}$) and model-predicted \Tclr ($T_\mathrm{clr}^\mathrm{model}$) trace each other, with $\mathrm{RMSD}=1.6$~K. However, $T_\mathrm{clr}^\mathrm{measured}$ has a smaller dynamic range than $T_\mathrm{clr}^\mathrm{model}$. At $\Tclr \lesssim (\gtrsim) 22.5$~K, the model tends to underestimate (overestimate) \Tclr. The difference at the highest-temperature end is possibly due to the scatter resulting from lack of measurements: although we have $\sim 16$ per cent of data points with $T_\mathrm{clr}^\mathrm{measured}>22.5$~K, only $\sim2$ per cent of the data have $T_\mathrm{clr}^\mathrm{measured}>25$~K. As mentioned in Section~\ref{sec:measurements:Tclr-SigmaSFR}, a larger sample focusing on high \Sigmasfr (i.e.\ high \Tclr) is necessary to draw a firm conclusion for the dust properties at the high-\Tclr end. The difference in the low-temperature part connects back to the underestimation of \Tclr in low-\Sigmasfr regions described in Section~\ref{sec:comparison:Tclr_SigmaSFR}. As shown in the left panel of Fig.~\ref{fig:Tclr-Tclr}, the threshold below which the model underestimates \Tclr is near $\Sigmasfr\sim10^{-2}~\SigmasfrUnit$, the same threshold we found in Section~\ref{sec:comparison:Tclr_SigmaSFR}. The $T_\mathrm{clr}^\mathrm{model}$ has a correlation coefficient of $0.97$ with $\log\Sigmasfr$, $\sim 0.09$ higher than the one of $T_\mathrm{clr}^\mathrm{measured}$, indicating a slight overestimation of the role of \Sigmasfr in dust heating in the model.
On the other hand, $T_\mathrm{clr}^\mathrm{model}$ has a correlation coefficient of $0.67$ with $\log\Sigmad$, $\sim 0.06$ higher than the one with $T_\mathrm{clr}^\mathrm{measured}$, indicating a possible underestimation of the strength of dust shielding.

\subsection{Implications for the Hirashita \& Chiang (2022) model}\label{sec:comparison:implication}

Through the various comparisons, we found that the analytical model proposed in \citetalias{Hirashita_Chiang22} successfully reproduced the first-order dependence of \Tclr on \Sigmasfr. The predicted \Tclr has correlation coefficients with \Sigmasfr and \Sigmad quite similar to the observations. The predicted \Tclr from \Sigmasfr and \Sigmad traces the observed \Tclr, with a RMSD of 1.6~K.

The prediction of higher \Tclr toward lower D/G at fixed \Sigmasfr is less obvious in the observations. The phenomena are only seen above $\Sigmasfr\sim10^{-2}~\SigmasfrUnit$ (or $\Tclr\sim22.5$~K; Fig.~\ref{fig:2kpc_SigmaSFR_T(250_100)_sim}). Below that, the model tends to underestimate \Tclr, and the observed \Tclr is not sensitive to D/G at fixed \Sigmasfr.
This suggests that there are physical mechanisms not modeled in the low-\Sigmasfr region, which will be discussed below. However, this caveat will likely not affect the high-redshift applications, which the model was originally designed for, as high-redshift targets usually have \Sigmasfr higher than the sample in this paper.

\begin{figure}
 \centering
 \includegraphics[width=0.99\columnwidth]{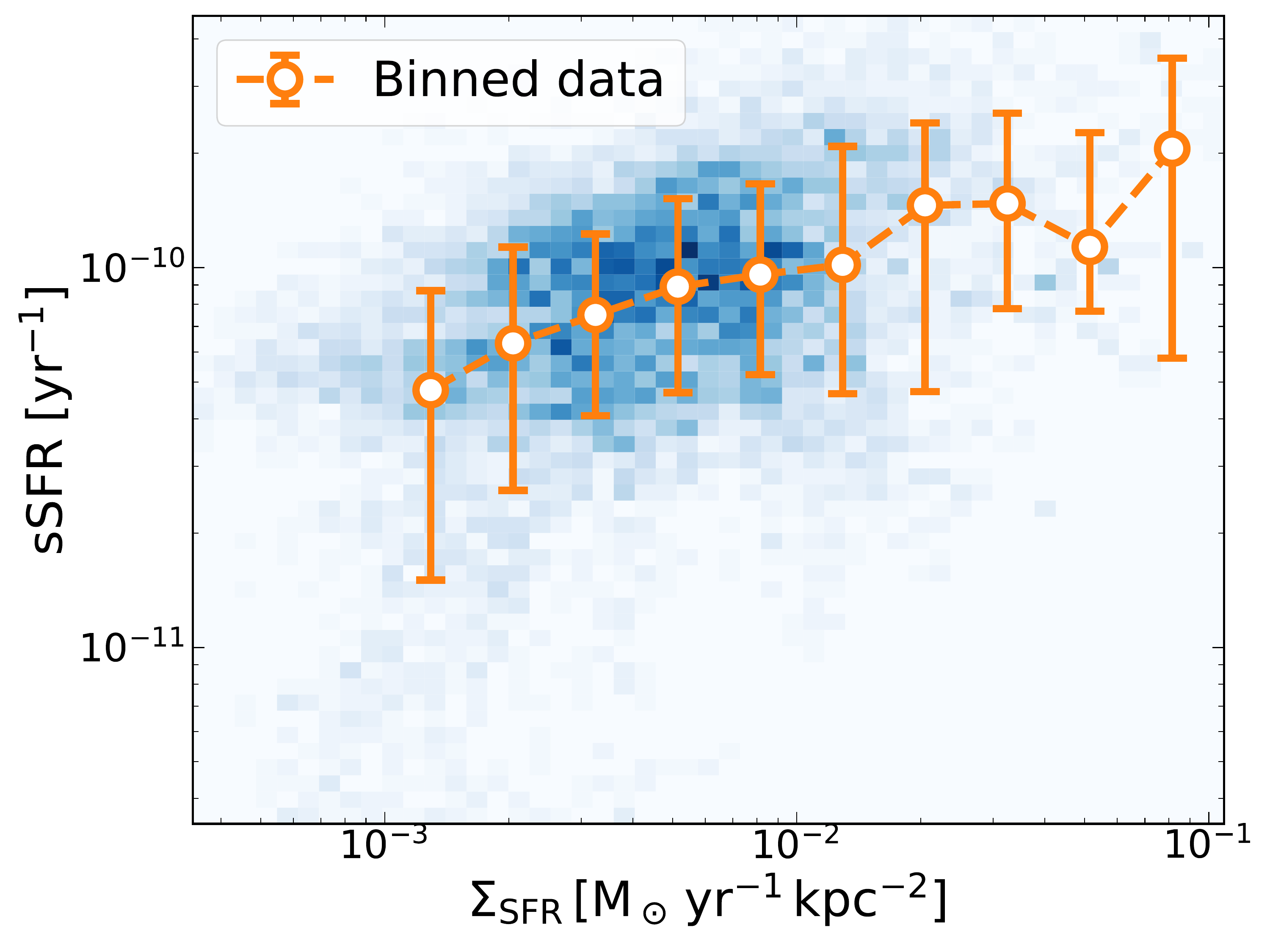}
    \caption{Relation between sSFR and \Sigmasfr in our sample. The points connected by the dashed line show the median of sSFR in each \Sigmasfr bin with the error bar showing the standard deviation, and the blue colour level shows the density of the pixel-by-pixel measurements on this diagram.
    }
    \label{fig:sSFR-SFR}
\end{figure}

We raise three possible explanations for the higher observed temperature at low-\Sigmasfr. The first possibility is that the contribution to dust heating from old stars might be more important than assumed in the \citetalias{Hirashita_Chiang22} model \citep{GROVES12,DeLooze14,Nersesian19}. For instance, \citet{Abdurro'uf22a} showed, using a spatially resolved analysis of nearby galaxies, that the energy contribution from old stars to dust heating increases as sSFR decreases \citep[see also][]{Boquien16,Leja19}. In Fig.~\ref{fig:sSFR-SFR}, we show the measured sSFR as a function of \Sigmasfr. We observe that the sSFR becomes lower towards lower \Sigmasfr. This trend means that the fraction of young stars is up to three times lower comparing to old stars, not emitting UV radiation, at the lowest \Sigmasfr bin. This observation suggests that the assumption of young-star-dominated heating is only applicable above certain threshold of \Sigmasfr, which is around $\Sigmasfr\sim 1\times10^{-2}~\SigmasfrUnit$.

The second explanation is the difference in the stellar ages set by the model and traced by observations. In the model, we assume that the SFR is constant over the past $\tau_\star=10^{10}$~yr. However, our FUV and IR indicators only trace stars with age up to $\sim 10^8$~yr \citep[see][and references therein]{KENNICUTT12}. If the SFR at age > $10^8$~yr is higher than the current SFR \citep[e.g.~the decaying SFR in][]{Nersesian19}, the constant-SFR model would underestimate the dust heating contributed from old ($>10^8$ yr) stellar populations and thus the dust temperature.

The above two possibilites can be tested by adding a stellar SED of an old stellar population to the \citetalias{Hirashita_Chiang22}  model. For the purpose of this test, we add a stellar SED generated by \textsc{starburst99} with an instantaneous burst occurring 1 Gyr ago, which is old enough not to contribute to the observed SFR, but young enough to contribute significantly to the dust temperature. Moreover, we choose the total stellar mass formed by this instantaneous burst two times larger than the original component (formed by a constant SFR over time), so that the sSFR is roughly three times lower. As a consequence, the dust temperature at $\Sigmasfr=10^{-3}$~\SigmasfrUnit is raised to 18 and 19 K for $\mathrm{D/G}=10^{-2}$ and $2\times 10^{-3}$, respectively (originally $\sim 16$~K). These temperatures are consistent with the observed values, thus supporting the contribution from old stellar population as a reason for the higher dust temperatures in the observations than in the \citetalias{Hirashita_Chiang22} model.
Since the old stellar population contributes to emission at longer wavelengths, its emission is less affected by dust extinction. This may also explain the small difference between the two cases of D/G values at low \Sigmasfr.

The third explanation we bring up is the spatial distribution of dust. In the \citetalias{Hirashita_Chiang22} model, we assume a uniform distribution of dust and stars in the plane of the galaxy disc, and adopt a thin-disc approximation. However, it is possible that in the outer disc, dust is less localized with young stars. This is supported by the residual shown in Fig.~\ref{fig:2kpc_SigmaSFR_SigmaDust}: in the low-\Sigmasfr region, the observation deviates toward higher \Sigmasfr per \Sigmad, in other words, lower \Sigmad at fixed \Sigmasfr. This trend effectively lowers down the dust opacity and could raise the dust temperature. \citet{Alton98}, \citet{Bianchi07}, \citet{Munoz-Mateos09} and \citet{Hunt15_dust} also showed that the exponential scale length of dust is longer (more extensive) than the one of stars. However, we remind the readers that there are also literature showing consistent spatial distribution between dust and FUV, e.g.\ \citet{Casasola17} showed that \Sigmad and $I_\mathrm{FUV}$ have almost identical mean scale lengths ($h\sim 0.4R_{25}$) in the DustPedia sample.

\section{Summary}\label{sec:summary}
In this work, we investigate the spatially resolved properties of dust temperature in nearby galaxies. We examine how the dust temperature is regulated by the surface densities of various quantities such as \Sigmasfr, \Sigmad, and \Sigmagas, with both observations and model predictions. To achieve this goal, we compile multi-wavelength observations of dust, stars and SFR in 46 nearby galaxies, and make a 2~kpc scale map of each component. We measure how dust temperature (the 250~\micron-to-100~\micron\ colour temperature) scales with local physical conditions, and then compare our measurements to the dust temperature model developed by \citetalias{Hirashita_Chiang22}. We find the following features for the measured dust properties:
\begin{itemize}
    \item The measured \Tclr correlates well with spatially resolved \Sigmasfr. The best-fit power law yields a RMSD of 0.81~K.
    \item None of the measured quantities has an overall strong correlation with the residual of the above power-law fitting, $\Delta$\Tclr. When analyzed in each \Sigmasfr bin, we find some trends:
    Both D/G and \Sigmad have stronger negative correlations with $\Delta$\Tclr at mid to high \Sigmasfr, meaning that the effect of lower \Tclr due to increased dust opacity is stronger at high-\Sigmasfr regions.
    \item We provide an empirical formula to predict \Sigmasfr with \Sigmad and D/G (equation~\ref{eq:KSDust_D/G}). This is equivalent to a KS law modified by a secondary dependence on D/G. 
\end{itemize}

Here is the summary of our key findings by comparing the measured dust temperature to the one predicted with the \citetalias{Hirashita_Chiang22} analytical dust temperature model:
\begin{itemize}
    \item Our observations show that \Tclr strongly correlates with \Sigmasfr and that at fixed \Sigmasfr, \Tclr increases as D/G decreases at $\Sigmasfr \gtrsim 1.0 \times 10^{-2}~\SigmasfrUnit$. These results are consistent with the \citetalias{Hirashita_Chiang22} model. The latter is interpreted as less dust shielding in a dust-poor environment.
    \item The \Tclr predicted from \Sigmasfr and \Sigmad by the \citetalias{Hirashita_Chiang22} model using the newly derived star formation relation is reasonably consistent with observations, with a RMSD of 1.6~K.  However, we observe a weaker dependence of \Tclr on \Sigmasfr with our measurements.
    \item At low \Sigmasfr ($\lesssim 10^{-2}$~\SigmasfrUnit), our observed \Tclr is higher than the prediction from the \citetalias{Hirashita_Chiang22} model, and it has no significant dependence on D/G. This is likely due to the contribution of dust heating from old stellar population and/or the variation of SFR within the past $10^{10}$~yr.
\end{itemize}

We conclude that the dust temperature reasonably scales with \Sigmasfr. At high-\Sigmasfr, the \Tclr at fixed \Sigmasfr decreases as D/G increases, which means that the \citetalias{Hirashita_Chiang22} model is consistent with observations at $\Sigmasfr \gtrsim 10^{-2}~\SigmasfrUnit$. Thus, we confirm that dust heating from young stars and shielding of radiation by dust regulate the dust temperature.
We also propose a prescription to predict \Sigmasfr from \Sigmagas and D/G, i.e.\ a D/G-modified KS law.
In the end, we succeed in obtaining a comprehensive understanding for the relations among dust temperature, dust content and star formation activity.

\section*{Acknowledgements}

We thank the anonymous referee for useful comments that helped to improve the quality of the manuscript.
IC thanks E. Schinnerer for useful discussions about this work.
IC and HH thank the National Science and Technology Council for support through grants 108-2112-M-001-007-MY3 and 111-2112-M-001-038-MY3, and the Academia Sinica for Investigator Award AS-IA-109-M02.
AS is supported by an NSF Astronomy and Astrophysics Postdoctoral Fellowship under award AST-1903834.
EWK acknowledges support from the Smithsonian Institution as a Submillimeter Array (SMA) Fellow and the Natural Sciences and Engineering Research Council of Canada (NSERC).
JS acknowledges support by NSERC through a Canadian Institute for Theoretical Astrophysics (CITA) National Fellowship.

This work uses observations made with ESA \textit{Herschel} Space Observatory. \textit{Herschel} is an ESA space observatory with science instruments provided by European-led Principal Investigator consortia and with important participation from NASA.

This paper makes use of the VLA data with project codes 14A-468, 14B-396, 16A-275 and 17A-073, which has been processed as part of the EveryTHINGS survey.
This paper makes use of the VLA data with legacy ID AU157, which has been processed in the PHANGS-VLA survey.
The National Radio Astronomy Observatory is a facility of the National Science Foundation operated under cooperative agreement by Associated Universities, Inc. 
This publication makes use of data products from the Wide-field Infrared Survey Explorer, which is a joint project of the University of California, Los Angeles, and the Jet Propulsion Laboratory/California Institute of Technology, funded by the National Aeronautics and Space Administration.

This paper makes use of the following ALMA data, which have been processed as part of the PHANGS-ALMA CO(2--1) survey: \linebreak
ADS/JAO.ALMA\#2012.1.00650.S, \linebreak 
ADS/JAO.ALMA\#2015.1.00782.S, \linebreak 
ADS/JAO.ALMA\#2018.1.01321.S, \linebreak 
ADS/JAO.ALMA\#2018.1.01651.S. \linebreak 
ALMA is a partnership of ESO (representing its member states), NSF (USA) and NINS (Japan), together with NRC (Canada), MOST and ASIAA (Taiwan), and KASI (Republic of Korea), in cooperation with the Republic of Chile. The Joint ALMA Observatory is operated by ESO, AUI/NRAO and NAOJ.

This research made use of Astropy,\footnote{http://www.astropy.org} a community-developed core Python package for Astronomy \citep{ASTROPY13,Astropy18,Astropy22}. 
This research has made use of NASA's Astrophysics Data System Bibliographic Services. 
We acknowledge the usage of the HyperLeda database (http://leda.univ-lyon1.fr). 
This research has made use of the NASA/IPAC Extragalactic Database (NED), which is funded by the National Aeronautics and Space Administration and operated by the California Institute of Technology.

\section*{Data Availability}
Data related to this publication and its figures are available on reasonable request
from the corresponding author.



\bibliographystyle{mnras}
\bibliography{references} 




\appendix

\section{Consistency between dust temperature tracers}\label{app:Tdust-Tracers}

\begin{figure}
    \centering
    \includegraphics[width=0.99\columnwidth]{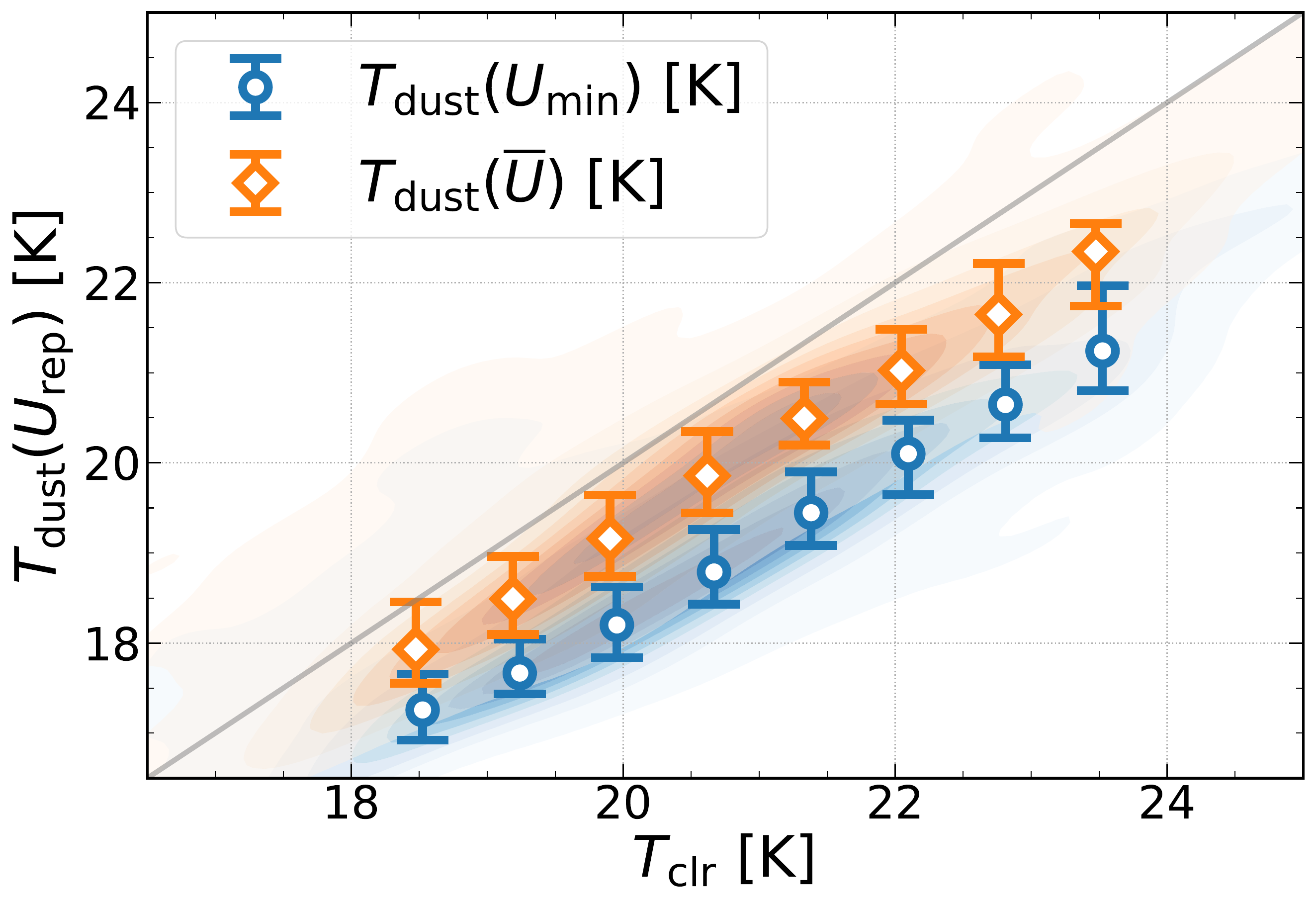}
    \caption{Comparison between the dust temperature tracers. The horizontal axis shows \Tclr, the FIR-based single dust temperature. The vertical axis presents $\Td(U_\mathrm{rep})$, the effective temperature converted from the ISRF distribution function.
    The blue shades show the density plot of pixel-by-pixel measurements of \Tumin, and the blue line shows the median in each \Tclr bin. The orange shades and line show the counterparts for \Tubar.}
    \label{fig:T_comparison}
\end{figure}

In our analysis, we use \Tclr as the fiducial tracer for dust temperature. There are other commonly used tracers for dust temperature that can be derived from dust SED fitting.
Broadly, there are two different approaches: (1) the single temperature derived from the modified blackbody model \citep{Schwartz1982,HILDEBRAND83_MBB}, which focuses on the emission in the FIR from large grains in radiative equilibrium and (2) the effective temperature derived from the distribution function of ISRF, which is a product of SED fitting with a physical dust model \citep[e.g.][]{DALE01,DRAINE07}. The latter usually includes the non-equilibrium thermal emission from small grains.
In this section, we briefly examine the consistency between these two approaches, i.e.\ the colour temperature (\Tclr) introduced in the main text, which focuses on the equilibrium emission in the FIR thus belongs to the first approach, and the ISRF strength from SED fitting results in $z$0MGS \citep[J. Chastenet et al. in preparation, fitting done with the][model]{DRAINE07}, which belongs to the second method.

In the \citet{DRAINE07} model, the scaled ISRF intensity is denoted as $U$, expressed in units of the MW diffuse ISRF at 10~kpc from \citet{Mathis83}. The ISRF contributing to dust heating is described by a combination of two components: a fraction of dust mass ($\gamma$) heated by a power-law distribution ($\propto U^{-\alpha}$) of starlight with $\Umin \leq U \leq U_\mathrm{max}$ \citep{DALE01}, and the rest ($1-\gamma$) of dust mass heated by $U=\Umin$, i.e.\ diffuse ISM radiation.

J. Chastenet et al. (in preparation) fix $\alpha=2$ and $U_\mathrm{max}=10^7$, and leaves $\gamma$ and \Umin as free parameters \citep[following][]{ANIANO12,ANIANO20}. According to \citet{DRAINE14}, we can convert a representative ISRF ($U_\mathrm{rep}$) to a representative dust temperature with:
\begin{equation}\label{eq:U2T}
    \frac{\Td(U_\mathrm{rep})}{1~\rm K} = 18 \times U_\mathrm{rep}^{1/6}~.
\end{equation}
There are two commonly used conventions for $U_\mathrm{rep}$. One is $U_\mathrm{rep}=\Umin$ by assuming that the majority of dust mass is heated by \Umin, i.e.\ $\gamma \ll 1$. This assumption is true almost throughout the entire M101 \citep{Chastenet21_M101} and regions with high dust luminosity surface density \citep{ANIANO12,Chastenet21_M101}. The other convention adopts the dust-mass-averaged ISRF, \Ubar, as $U_\mathrm{rep}$. The functional form of \Ubar is:
\begin{equation}\label{eq:Ubar}
    \Ubar = (1-\gamma)\Umin + \gamma\Umin\frac{\ln(U_\mathrm{max}/\Umin)}{1-(\Umin/U_\mathrm{max})}~~~\text{for $\alpha=2$}.
\end{equation}

With that said, the three \Td tracers we will compare here are: \Tclr, $\Td(\Umin)$ and $\Td(\Ubar)$ (equations~\ref{eq:Tclr}, \ref{eq:U2T} and \ref{eq:Ubar}). As shown in Fig.~\ref{fig:T_comparison}, the three tracers are roughly parallel to each other in the temperature range of interest, suggesting that these temperature tracers reasonably trace each other with a possible systematic offset in the calibration between \Tclr and the representative \Td converted from the ISRF strength. \Tclr is systematically $\sim 0.8~\rm K$ and $\sim 1.8~\rm K$ higher than \Tubar and \Tumin, respectively. The fact that \Tubar is consistently higher than \Tumin suggests that the hotter dust component, which is represented by the power-law component in the \citet{DRAINE07} formulation, is important in the model. The $\sim 0.4~\rm K$ ($\sim 2\%$ in \Td, or $\sim 0.05$~dex in $U$) offset between \Tclr and \Tubar is roughly 2.5 times the typical uncertainty in \Ubar from J. Chastenet et al. (in preparation, $\sigma\sim 0.02~\mathrm{dex}$), which indicates that this offset is significant.

\section{Robustness under the adopted CO-to-H\texorpdfstring{$_2$}{2} conversion factors}\label{app:other_aCO}

As we mentioned in Section~\ref{sec:method:physical}, how the CO-to-H$_2$ conversion factor, \aco, depends on local conditions is still an active field of study. In our main analysis, we calculate \aco with the prescription suggested by \citet{BOLATTO13}, because it is one of the few prescriptions that formulate both the CO-dark molecular gas and the decrease of \aco in galaxy centres, where the latter has been proved to be necessary to obtain reasonable values for D/G and D/M \citep{Chiang21}. Meanwhile, \citet{Sun20} suggested that the exponential dependence on metallicity in the \citet{BOLATTO13} prescription might overestimate \aco in low-metallicity regions. Besides an exponential dependence, it is a commonly used strategy to formulate \aco as a power-law of metallicity \citep[e.g.][]{GloverMacLow11,SCHRUBA12,HUNT15}. In this section, we examine how our findings might change with different \aco prescriptions. Specifically, we will examine (1) how the correlation between $\Delta$\Tclr and D/G changes with \aco (Sections~\ref{sec:measurements:Tclr-SigmaSFR} and \ref{sec:comparison:Tclr_SigmaSFR}), and (2) whether the D/G modification to the KS law still makes sense (Section~\ref{sec:measure:KS law}).

\begin{figure}
    \centering
    \includegraphics[width=0.99\columnwidth]{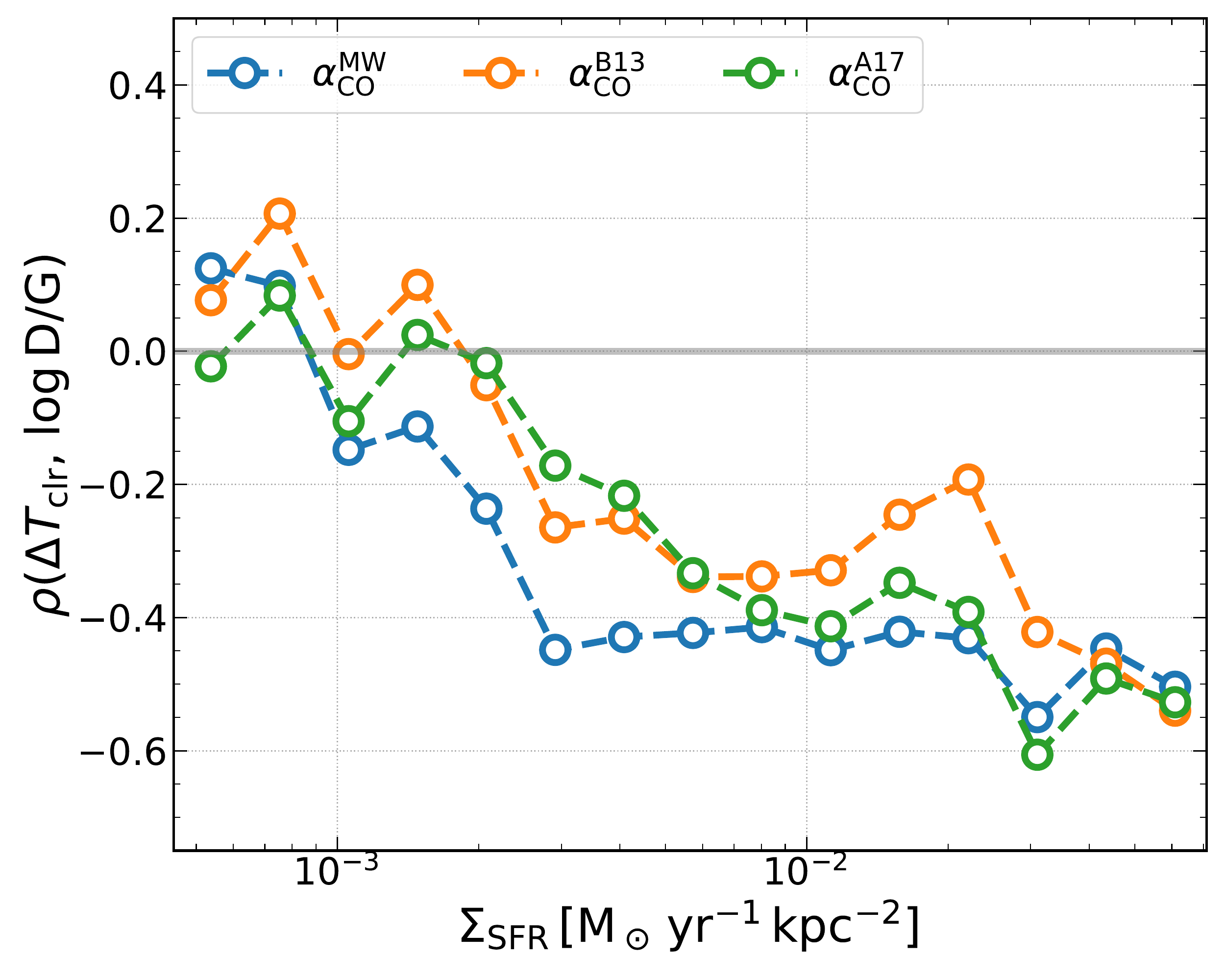}
    \caption{Pearson's correlation coefficient ($\rho$) between $\Delta$\Tclr and $\log\mathrm{(D/G)}$ calculated with different \aco prescriptions.}
    \label{fig:dT-corr-aCO}
\end{figure}

We adopt three \aco prescriptions here: a constant \aco, an \aco that depends on metallicity exponentially, and an \aco that depends on metallicity with a power law. For the constant one, we adopt the \aco in typical giant molecular clouds (GMCs) in the MW disc, $\acoMW = 4.35~\acoUnit$ \citep{SOLOMON87_1987ApJ...319..730S,STRONG96_1996AA...308L..21S,ABDO10_2010ApJ...710..133A}. We use the \citet{BOLATTO13} prescription (equation~\ref{eq:acoBolatto}) for the exponential one. Lastly, we adopt the power-law prescription suggested by \citet{ACCURSO17}, who used the [\textsc{C~ii}] to CO $J=1\to0$ ratio to constrain \aco and obtained the following metallicity-dependent conversion factor:
\begin{equation}\label{eq:A17 aCO}
    \frac{\acoAccurso}{1~\acoUnit} = 4.35\,Z'^{-1.6}.
\end{equation}
Note that we neglect the dependence on the offset from the main sequence in the \citet{ACCURSO17} formulation, following the approach in \citet{Sun20}.

In addition to \aco, we also examine how the variation in the CO line ratio, $R_{21}$ (equation~\ref{eq:CO_R21}), might affect our results. We adopt the SFR-dependent formula proposed in \citet[][see their Table 5]{Leroy21_CO_Line_Ratios}, that is:
\begin{equation}\label{eq:R21_variation}
    \log\bigg(\frac{R_{21}}{\langle R_{21}\rangle_\mathrm{gal}}\bigg) = 0.129\log\bigg(\frac{\Sigmasfr}{\langle\Sigmasfr\rangle_\mathrm{gal}}\bigg) + 0.019,
\end{equation}
where the $\langle\rangle_\mathrm{gal}$ symbol represents the average value in the disc of each galaxy. For simplicity, we use $
\langle R_{21}\rangle_\mathrm{gal}=0.65$ for all galaxies, and we adopt the geometric mean of the SFR surface densities for all pixels within a galaxy for $\langle\Sigmasfr\rangle_\mathrm{gal}$. Following the suggestion in \citet{Leroy21_CO_Line_Ratios}, if the derived $R_{21}$ is greater than 1.0, we directly set it to 1.0.

We first re-calculate the correlation between $\Delta$\Tclr (Section~\ref{sec:measurements:Tclr-SigmaSFR}) and $\log(\mathrm{D/G})$ with the three \aco prescriptions.
The correlation coefficients are calculated for each \Sigmasfr bin. Compared to Fig.~\ref{fig:dT-corr}, we only keep D/G because it is the only quantity affected by the choice of \aco. As shown in Fig.~\ref{fig:dT-corr-aCO}, the correlation coefficients as a function of \Sigmasfr have similar behaviours for the three \aco prescriptions. They all show poor correlation at low \Sigmasfr, and the correlations become stronger toward high \Sigmasfr. The two metallicity-dependent \aco prescriptions have similar transition \Sigmasfr from poor to stronger correlation. Thus, as long as we use a metallicity-dependent \aco, we expect \Tclr to have poor correlation with D/G at low \Sigmasfr. The correlations between $\Delta$\Tclr and $\log(\mathrm{D/G})$ calculated with varying $R_{21}$ (equation~\ref{eq:R21_variation}) does not deviate significantly from the constant $R_{21}$ results.

\begin{table}
\centering
\caption{The RMSD (in dex) of the power-law fittings to \Sigmasfr using different sets of variables for the three \aco prescriptions. For the fitting with two variables (\Sigmad and D/G), we examine two cases for $R_{21}$. See Section~\ref{sec:measure:KS law} for fitting formulation.}
\label{tab:RMSE_aCO}
\begin{tabular}{lccc}
\toprule
Formulation & \acoMW & \acoBolatto & \acoAccurso \\
\midrule
\Sigmasfr(\Sigmad) & 0.27 & \nodata & \nodata \\
\Sigmasfr(\Sigmad, D/G), constant $R_{21}$ & 0.20 & 0.22 & 0.22 \\
\Sigmasfr(\Sigmad, D/G), varying $R_{21}$ & 0.22 & 0.24 & 0.24 \\
\bottomrule
\end{tabular}
\begin{tablenotes}
    \small
    \item {\bf Note:} There is no \aco involved in the \Sigmasfr(\Sigmad) fitting.
\end{tablenotes}
\end{table}

Next, we examine the power-law fittings to \Sigmasfr similar to what we have done in Section~\ref{sec:measure:KS law} with \acoBolatto. In Table~\ref{tab:RMSE_aCO}, we present the RMSD of all the power-law fittings in this test. We test three formulations: the \Sigmasfr(\Sigmad) and \Sigmasfr(\Sigmad, D/G) with constant $R_{21}$, which we have performed in Section~\ref{sec:measure:KS law}, and \Sigmasfr(\Sigmad, D/G) with varying $R_{21}$ in addition. 
As shown in Table~\ref{tab:RMSE_aCO}, all the three \aco prescriptions show improved RMSD when moving from single-variable power law to the two-variable power law (constant $R_{21}$), indicating that the D/G-modification to the standard KS law (equation~\ref{eq:KS_D/G}) is supported robustly against the change of the \aco prescriptions. Meanwhile, the two-variable power law with varying $R_{21}$ has systematically larger RMSD than the constant $R_{21}$ ones, showing the variation of $R_{21}$ does not improve the fitting.

When we interpret the two-variable power law as a D/G-modification to the standard KS law (equation~\ref{eq:KS_D/G}) with different \aco prescriptions, we find the coefficient for the $\log\Sigmagas$ term (corresponding to the $N$ value in the standard KS law) as 1.35 for \acoAccurso and 1.39 for \acoMW, both lower than the 1.49 for \acoBolatto. Meanwhile, we have similar coefficients for the $\log(\mathrm{D/G})$ term for the metallicity-dependent prescriptions: 0.72 for \acoAccurso and 0.76 for \acoBolatto. On the other hand, the coefficient for the constant \acoMW is 0.48, which yields a weaker correction than the metallicity-dependent \aco prescriptions.

\bsp	
\label{lastpage}
\end{document}